\documentclass[universe,article,submit,pdftex,moreauthors]{Definitions/mdpi} 
\usepackage{bigints}
\usepackage{xcolor}
\usepackage{comment}
\usepackage{graphicx}	
\usepackage{amsmath}	
\usepackage{amssymb}	
\usepackage{soul}
\usepackage{caption}
\usepackage{subcaption}
\usepackage{appendix}
\usepackage{mwe}
\usepackage{ulem}
\usepackage{nomencl}
\makenomenclature
\usepackage{float}

\firstpage{1} 
\pubvolume{1}
\issuenum{1}
\articlenumber{0}
\pubyear{2024}
\copyrightyear{2024}
\datereceived{ } 
\daterevised{ } 
\dateaccepted{ } 
\datepublished{ } 
\hreflink{https://doi.org/} 

\Title{Constraining the Initial-Mass Function via Stellar Transients}

\TitleCitation{Constraints on the IMF from stellar transients}

\Author{Francesco Gabrielli$^{1,2}$\orcidF{}, Lumen Boco$^{1,2,3}$\orcidL{}, Giancarlo Ghirlanda$^{4,5}$\orcidG{}, Om Sharan Salafia$^{4,5}$\orcidO{}, Ruben Salvaterra$^{6}$\orcidR{}, Mario Spera$^{1,2,7}$\orcidM{}, Andrea Lapi$^{1,2,3,8}$\orcidA{}}
\address{
$^{1}$ \quad Scuola Internazionale Superiore Studi Avanzati (SISSA), Physics Area, Via Bonomea 265, 34136 Trieste, Italy\\
$^{2}$ \quad INFN-Sezione di Trieste, Via Valerio 2, 34127 Trieste,  Italy\\
$^{3}$ \quad IFPU-Institute for Fundamental Physics of the Universe, Via Beirut 2, 34014 Trieste, Italy\\ 
$^{4}$ \quad INAF - Osservatorio Astronomico di Brera, Via E Bianchi 46, I-23807 Merate (LC), Italy\\
$^{5}$ \quad INFN - Sezione di Milano-Bicocca, piazza della Scienza 3, I-20126 Milano (MI), Italy\\
$^{6}$ \quad INAF – Istituto di Astrofisica Spaziale e Fisica Cosmica di Milano, Via A. Corti 12, 20133 Milano, Italy\\
$^{7}$ \quad Istituto Nazionale di Astrofisica - Osservatorio Astronomico di Roma, Via Frascati 33, I-00040, Monteporzio Catone, Italy\\
$^{8}$ \quad IRA-INAF, Via Gobetti 101, 40129 Bologna, Italy
}

\AuthorCitation{F. Gabrielli et al.}

\abstract{The stellar initial-mass function (IMF) represents a fundamental quantity in astrophysics and cosmology, describing the mass distribution of stars from low mass all the way up to massive and very-massive stars. It is intimately linked to a wide variety of topics, including stellar and binary evolution, galaxy evolution, chemical enrichment, and cosmological reionization. Nonetheless, the IMF still remains highly uncertain. In this work, we aim at determining the IMF with a novel approach based on the observed rates of transients of stellar origin. We parametrize the IMF with a simple, but flexible, Larson shape, and insert it into a parametric model for the cosmic UV luminosity density, local stellar mass density, type Ia supernova (SN Ia), core-collapse supernova (CCSN), and long gamma-ray burst (LGRB) rates as function of redshift. We constrain our free parameters by matching the model predictions to a set of empirical determinations for the corresponding quantities, via a Bayesian Markov-Chain Monte Carlo method. Remarkably, we are able to provide an independent IMF determination, with characteristic mass $m_c=0.10^{+0.24}_{-0.08}\:M_{\odot}$, and high-mass slope $\xi=-2.53^{+0.24}_{-0.27}$, that is in accordance with the widely-used IMF parameterizations (e.g. Salpeter, Kroupa, Chabrier). Moreover, the adoption of an up-to-date recipe for the cosmic metallicity evolution, allows us to constrain the maximum metallicity of LGRB progenitors to $Z_{max}=0.12^{+0.29}_{-0.05}\:Z_{\odot}$. We also find what progenitor fraction actually leads to SN Ia or LGRB emission (e.g. due to binary interaction or jet-launching conditions), put constraints on the CCSN and LGRB progenitor mass ranges, and test the IMF universality. These results show the potential of this kind of approach for studying the IMF, its putative evolution with galactic environment and cosmic history, and the properties of SN Ia, CCSN and LGRB progenitors, especially considering the wealth of data incoming in the future.}


\begin{document}

\section{Introduction}

The stellar initial-mass function (IMF) describes the mass distribution of stars on the zero-age main sequence (ZAMS). It plays a pivotal role in many different topics of astrophysics and cosmology, spanning from stellar and binary evolution, to galaxy evolution and cosmological reionization. Past and current empirical determinations of the IMF have been relying either on the observation of stellar populations inside our Galaxy and in the Local Group, or on the study of the suggested correspondence between the IMF and the mass function of observed pre-stellar dense cores (see e.g. the recent review by \cite{Hennebelle_2024} and references therein). Regarding the first method, the IMF can be inferred either from direct star counts, in the case of resolved stellar populations, or via careful modeling of e.g. galaxy spectra or gravitational lensing in the case of unresolved stellar populations (\cite{Hennebelle_2024} and references therein). Based on these observational results, a number of parametrizations have been introduced to model the IMF, the most common being those by \cite{Salpeter_1955,Scalo_1986,Kroupa_1993,Scalo_1998,Kroupa_2001,Chabrier_2003,Kroupa_2013,Kroupa_2021}. However, as pointed out e.g. in \cite{Weisz_2012,Wainer_2024}, a large part of IMF measurements underestimate the scatter observed in the IMF slopes, as well as the difficulty in disentangling systematics from physical IMF environmental dependencies. These limitations also threaten the common assumption of an IMF universality, with a number of works suggesting the IMF dependence on e.g. redshift, metallicity and/or star formation rate (SFR) (e.g. \cite{Strolger_2015,Stanway_2019,Ziegler_2022,Li_2023,Martin-Navarro_2023,Maksymowicz-Maciata_2024,Wainer_2024}). On the theory side, numerous works have been trying to explain the physical origin of the IMF characteristic mass, identifying its peak, and high-mass slope (\cite{Hennebelle_2024} and references therein). Despite all these efforts, many fundamental uncertainties still exist around the properties of the IMF and its physical nature.\\

The IMF plays a key role in the study of stellar transients, since it describes the mass distribution of their progenitors. In particular, it is crucial for computing the cosmic rates of type Ia supernovae (SNe Ia) and core-collapse supernovae (CCSNe). Moreover, the IMF critically determines the amount of UV radiation emitted by stars in galaxies.\\

Young massive stars are responsible for the emission of ionizing radiation in the UV band. The corresponding UV luminosity density can be computed by integrating UV luminosity functions measured with galaxy surveys (e.g. \cite{Bouwens_2021,Bouwens_2023,Donnan_2023,Harikane_2023}). Since UV light traces the abundance of young massive stars, it is intimately linked to the star formation rate (SFR). Consequently, the cosmic UV luminosity density, $\rho_{UV}(z)$, can be converted to a cosmic star formation rate density (SFRD), $\rho_{SFR}(z)$, which indicates the amount of star-forming mass available at redshift $z$ per unit time and comoving volume. This conversion is recurrently performed using a factor $k_{UV}$, expressing the efficiency in the production of UV photons from stars. The value of $k_{UV}$ clearly depends on the IMF, as the latter determines the relative number of massive stars compared to low-mass ones (\cite{Kennicutt_2012,Foreman_Mackey_2013,Madau_2014,Cai_2014,Robertson_2015,Finkelstein_2019}).\\

SNe Ia are thermonuclear transients arising in binary systems composed of at least one white dwarf (WD), from progenitor stars of at least $2-3\:M_{\odot}$ (e.g. \cite{Madau_1998,Maoz_2014}). CCSNe, instead, represent the endpoint of massive star evolution above $\sim 8\:M_{\odot}$, and are believed to arise from the collapse of the more external layers of the star onto the iron core, and the subsequent, neutrino-driven explosion (e.g. \cite{Bethe_1985,Burrows_1995,Janka_1996,Burrows_2019,Aguilera_Dena_2023}). Great uncertainties revolve around the upper limit of CCSN progenitors, which might vary from $\sim$ 20-25 $M_{\odot}$ all the way up to 100 $M_{\odot}$ (e.g. \cite{Heger_2002,Heger_2003,Smartt_2015}). Indeed, stars inside this mass range are thought to experience fall-back of the ejected material onto the core, giving rise to a faint, "failed" SN, and collapse into a BH (e.g. \cite{Timmes_1996,Heger_2003,Zhang_2008,Smartt_2015,Coughlin_2018,Antoni_2023}). Above $\sim$100 $M_{\odot}$, stars develop pair-instability inside their core, and end their lives before completing all nuclear burning stages (\cite{Heger_2002,Heger_2003}). Given the dependence of SNe Ia and CCSNe on the progenitor mass, the IMF explicitly enters in setting their rates.\\

There exists another class of stellar transients that, differently from SNe Ia and CCSNe, exhibit a crucial dependence on metallicity, i.e. long gamma-ray bursts (LGRBs). According to the "collapsar" scenario, LGRBs arise from the collapse of massive stars, and the formation of an accretion disk around the BH remnant. If the disk is sufficiently massive, and the BH sufficiently spinning, a jet can be launched perpendicularly to the disk, powering the LGRB emission (\cite{Meszaros_2006}). Host galaxy studies and simulations show a LGRB preference for low-metallicity environments, suggesting the existence of a maximum metallicity above which stars cannot produce these transients (e.g. \cite{Heger_2003}). A number of LGRBs have been observed in association to type Ib/c SNe, suggesting a shared progenitor (e.g. \cite{Iwamoto_1998,Nomoto_2004,Woosley_2006,Mazzali_2006,Mazzali_2008,Valenti_2008,Hjorth_2011}). This might indicate a preferential origin in binary systems, where the progenitor can be stripped of its envelope by means of the companion. All in all, the properties of LGRB progenitors, and especially their mass range, are still unknown. The joint study of LGRBs and the cosmic metallicity ($Z$) evolution holds the potential to uncover the elusive progenitors of LGRBs, and increase our understanding of galaxy evolution.\\

As explained above, the IMF is deeply intertwined with both stellar transients and galaxy UV luminosity. Due to the realization of this \textit{fil rouge}, in this paper we are motivated to attempt at inferring the IMF with a novel method, based on combining observational constraints related to these quantities, in a way that is completely independent from previous IMF determinations.\\

Specifically, we build a parametric model for the cosmic UV luminosity density, local stellar mass density (SMD), SN Ia, CCSN, and LGRB rates as function of redshift, and constrain the model parameters in order to match the observational determinations available for these quantities. We do this by employing a Bayesian Markov-Chain Monte Carlo method (MCMC). Our approach also allows us to constrain the properties of SN Ia and CCSN progenitors. Moreover, the adoption of an up-to-date recipe for the cosmic metallicity evolution, enables us to study the properties of LGRB progenitors. Finally, we explore what the set of observational constraints adopted here can tell us about the putative IMF evolution with redshift, and discuss how variations in our model assumptions can alter our results.\\

This work is structured as follows. In Section \ref{sec:methods}, we describe how we compute the different quantities presented above, and introduce the free parameters of our model. Then we explain the basics of our MCMC, and present our results in Section \ref{sec:results}. We discuss our findings in Section \ref{sec:discussion}, as well as some variations in the model assumptions, and finally draw our conclusions in Section \ref{sec:conclusions}. Throughout this work we assume a flat $\Lambda$CDM cosmology, with parameters $H_0=70\:km\:s^{-1}\:Mpc^{-1}$, $\Omega_{M}=0.3$. We adopt the value of $Z_{\odot}=0.0153$ for the Solar metallicity (\cite{Caffau_2010}), and $12+\log$(O/H)$_{\odot}=8.76$ for the Solar oxygen abundance. Unless specified otherwise, we consider absolute metallicities. 

\section{Methods}\label{sec:methods}

We follow a parametric, semi-empirical approach in order to model the cosmic UV luminosity density, $\rho_{UV}(z)$, the local SMD, $\rho_{\star,0}$, and the SN Ia, CCSN and LGRB rates as a function of redshift, $R_{Ia}(z),\:R_{CC}(z),\:R_{LGRB}(z)$. We then employ a Bayesian MCMC method in order to match our model predictions to the data. This allows us to keep the number of physical assumptions to the minimum, and obtain an estimate of the model parameters that is directly informed by observations.\\

Throughout this work, we assume an IMF with Larson shape (\cite{Larson_1998}),
\begin{equation}
\phi(m)\propto m^{\xi}\:e^{-m_c/m},
\end{equation}
with free parameters $m_c$, the characteristic mass, and $\xi$, the slope.

\subsection{UV luminosity density}

The UV luminosity density, $\rho_{UV}(z)$, represents the energy in UV photons present at a given redshift per unit time and comoving volume, typically expressed in units of $erg\:s^{-1}\:Mpc^{-3}$. We adopt a standard, \cite{Madau_2014} functional form,
\begin{equation}
\rho_{UV}(z)=\rho_{UV,0}\frac{(1+z)^\alpha}{[1+(1+z)/\gamma]^\beta},
\end{equation}
where the normalization $\rho_{UV,0}$ is the local UV luminosity density. Based on the assumed IMF, one can convert this quantity to a SFRD, via the following equation:
\begin{equation}\label{eq:rho_sfr_uv}
\rho_{SFR}=\rho_{UV}/k_{UV}(m_c,\xi),
\end{equation}
where $k_{UV}$ is the UV photons production efficiency, in units of $erg\:s^{-1}\:Hz^{-1}\:M_{\odot}^{-1}\:yr$. We compute it by employing the \texttt{PARSEC} stellar evolution code (\cite{Bressan_2012,Chen_2014,Chen_2015,Tang_2014,Costa_2019,Costa_2021,Nguyen_2022,Goswami_2021,Goswami_2022}). In particular, we use stellar evolution tracks to compute the mean production rate of UV photons, and then average the result over the IMF to obtain $k_{UV}$. For simplicity, we adopt fixed values for the age and metallicity of the stellar population, $10^9\:yr$ and $Z=0.02$, respectively. We check that varying the stellar age to $10^8\:yr$ and $10^{10}\:yr$, and the metallicity to $Z=0.002$ and $0.0002$, has a negligible effect on the results.

In order to fit $\rho_{UV}$, we consider a set of the most recent SFRD determinations from UV, IR and submm surveys, including JWST (\cite{Gruppioni_2013,Rowan-Robinson_2016,Dunlop_2016,Novak_2017,Liu_2018,Oesch_2018,Bhatawdekar_2019,Gruppioni_2020,Bouwens_2023,Donnan_2023,Harikane_2023}). We convert from $\rho_{SFR}$ to $\rho_{UV}$ via Equation \ref{eq:rho_sfr_uv}, according to the IMF used in these works. We decide to restrict to data at $z<9$ since, as shown below, the observational constraints on the stellar transient rates employed in this work do not exceed this redshift (in particular those related to the LGRB rate). This choice also allows us to avoid the uncertainties in the SFRD data at higher redshifts.

\subsection{Local stellar mass density}\label{sec:smd_method}

By integrating the SFRD from $z=\infty$ down to a given redshift $z$, one can compute the total amount of star forming mass accumulated throughout the whole history of the Universe down to that redshift, per unit comoving volume, i.e. the SMD (see e.g. \cite{Madau_2014}):
\begin{equation}\label{eq:rho_star}
\rho_{\star}(z)=(1-R) \int_{\infty}^{z} \rho_{SFR}(z)\:\frac{dt}{dz}\:dz 
\end{equation}
$R$ accounts for the fraction of stellar mass returned into the ISM or IGM during stellar evolution, via winds or SN explosions. We compute it following \cite{Madau_2014}, under the assumption that stars above $1\:M_{\odot}$ lose mass the moment they are born ("instantenous recycling approximation"):
\begin{equation}
R=\int_{1\:M_{\odot}}^{300\:M_{\odot}} dm\:(m-\omega_m)\:\phi(m) ,
\end{equation}
where $\phi(m)$ is the IMF, and the upper limit of integration is the maximum stellar mass considered in this work. $\omega_m$ is the remnant mass, that we compute using the initial-final mass relation (IFMR) from \cite{Spera_2017} for $m>8\:M_{\odot}$, and that from \cite{Cummings_2018,Marigo_2020,Marigo_2022} for $m<8\:M_{\odot}$. In particular, we consider the \cite{Spera_2017} result for $Z=0.02$. We find that employing the IFMR for $Z=0.002$ and $Z=0.0002$ has a negligible effect on the results. The adopted IFMR for $m<8\:M_{\odot}$ is computed at $Z=0.014$. As discussed e.g. by \cite{Cummings_2018}, the IFMR is not expected to exhibit a crucial dependence on metallicity in this mass range.

In order to obtain an observational constraint on the local SMD, we consider a recent determination of the galaxy stellar mass functions (GSMFs) by \cite{Weaver_2023}, based on the COSMOS2020 galaxy catalogue. We focus on the results obtained in the lowest redshift bin, $z\sim 0.2-0.5$. GSMFs represent the number of galaxies per unit stellar mass and comoving volume, $\Phi(M_{\star})=d^{2}N/dM_{\star}\:dV$. By integrating over stellar mass, one can compute the total mass in stars per unit comoving volume, i.e. the SMD:
\begin{equation}\label{eq:rho_star_obs}
\rho_{\star,0}\equiv\rho_{\star}(z\sim 0.35)=\left( \int dM_{\star}\:M_{\star}\:\Phi(M_{\star}) \right)\:\frac{\bigl \langle M_{\star}/L_{UV} \bigr \rangle}{\bigl \langle M_{\star}/L_{UV} \bigr \rangle_{Ch}},
\end{equation}
where $z\sim 0.35$ is the mean redshift in the considered bin. The factors on the right of the integral are the stellar mass-to-light ratios, which we use to convert from a Chabrier IMF used in \cite{Weaver_2023}, to the IMF adopted here (as done e.g. in \cite{Madau_2014}). We compute these quantities via the \texttt{PARSEC} stellar evolution tracks. Finally, we compare this observational constraint with the SMD from Equation \ref{eq:rho_star}, computed at $z=0.35$.\\

\subsection{Core-collapse supernova rate}

Regarding the CCSN rate, we start by computing the number of CCSN produced per unit star forming mass, according to the following equation:
\begin{equation}
\frac{dn_{CC}}{dM_{\star}}=\int_{M_{CC}^{low}}^{M_{CC}^{up}} dm\:\phi(m) ,
\end{equation}
where we integrate the IMF over the mass range of CCSN progenitors. The minimum stellar mass to produce a CCSN, $M_{CC}^{low}$, is expected to be around $8\:M_{\odot}$ (e.g. \cite{Heger_2003,Smartt_2015}). On the other hand, the upper limit $M_{CC}^{up}$ is highly uncertain (e.g. \cite{Heger_2002,Heger_2003,Smartt_2015}). Based on these considerations, we decide to fix $M_{CC}^{low}=8\:M_{\odot}$, and keep $M_{CC}^{up}$ as a free parameter. Finally, we compute the CCSN rate as a function of redshift as:
\begin{equation}\label{eq:CCSN_rate}
R_{CC}(z)=\frac{dn_{CC}}{dM_{\star}}\times\rho_{SFR}(z)
\end{equation}
In doing so, we assume no delay between star formation and CCSN explosion, since the corresponding timescale ($\lesssim 100\:Myr$) is negligible on cosmological scales.\\

We compare to a set of observational determinations of the CCSN rate from the literature, spanning a redshift range up to $z\sim 2.3$ (\cite{Cappellaro_1999,Cappellaro_2005,Botticella_2007,Bazin_2009,Graur_2011,Li_2011,Dahlen_2012,Mattila_2012,Melinder_2012,Taylor_2014,Cappellaro_2015,Graur_2015,Strolger_2015,Petrushevska_2016,Frohmaier_2020}). This set comprises, to the best of our knowledge, all existing data available for this quantity. All rates are derived for $H_0=70\:km\:s^{-1}\:Mpc^{-3}$, the same value adopted in this work.

\subsection{Type Ia supernova rate}
SN Ia are thermonuclear explosions of WDs in binary systems, triggered either by accretion from a companion star ("single-degenerate" scenario, SD), or by merger with another WD ("double-degenerate", DD). We compute the number of SN Ia produced per unit star forming mass as
\begin{equation}\label{eq:SNIa_rate}
\frac{dn_{Ia}}{dM_{\star}}=N_{Ia}\:\int_{M_{Ia}^{low}}^{M_{Ia}^{up}} dm\:\phi(m)
\end{equation}
We agnostically account for all effects of binary evolution with the normalization factor $N_{Ia}$, similarly as done e.g. in \cite{Madau_1998,Kobayashi_2000a,Strolger_2004,Dahlen_2004,Valiante_2009,Melinder_2012,Cappellaro_2015,Petrecca_2024,Palicio_2024}. $N_{Ia}$ also includes the binary fraction, $f_{bin}$, i.e. the fraction of stars lying in a binary with respect to the whole stellar population. Then we integrate the IMF over the expected mass range of SN Ia progenitors. WD progenitors lie at masses starting from $\sim 0.7-0.8\:M_{\odot}$, i.e. the minimum mass for a star to complete the MS, and finally evolve into a WD, on a timescale less than the age of the Universe. However, it is believed that, in order for the WD to give rise to a SN Ia, the progenitor star must be at least of $2-3\:M_{\odot}$ (e.g. \cite{Madau_1998,Maoz_2014}). On the other hand, there is overall consensus placing the upper limit of SN Ia progenitors to $\sim 8\:M_{\odot}$, at the onset of core collapse. For these reasons, we fix $M_{Ia}^{low}=3\:M_{\odot}$, and $M_{Ia}^{up}\equiv M_{CC}^{low}=8\:M_{\odot}$. We check that moving $M_{Ia}^{low}$ to $2\:M_{\odot}$ does not affect our results significantly.
As in the case of CCSNe, we do not consider any dependence on metallicity, a standard assumption (but see e.g. \cite{Kobayashi_2000b,Toonen_2012,Kistler_2013,Claeys_2014}, where this assumption is challenged).

Since, differently from CCSNe, the timescales for SN Ia explosion after the birth of the progenitor star are not negligible, we compute the SN Ia rate by convolving the SFRD with a delay time distribution (DTD), $\Psi(t)$, following \cite{Maoz_2014}:
\begin{equation}
R_{Ia}(z)=\frac{dn_{Ia}}{dM_{\star}}\times \int_{40\:Myr}^{t(z)} d\tau\:\rho_{SFR}(t-\tau)\:\Psi(\tau) ,
\end{equation}
where $t(z)$ is the age of the Universe at redshift $z$, and we choose the lower cut of $40\:Myr$ as the delay time associated to stars of $\sim 8-10\:M_{\odot}$. We adopt a DTD of the form $\propto t^{-1}$, following \cite{Maoz_2012,Maoz_2014} and references therein.\\

We match this theoretical SN Ia rate to a set of observational determinations from the literature, reaching up to $z\sim 2.3$ (\cite{Pain_1996,Cappellaro_1999,Hardin_2000,Pain_2002,Madgwick_2003,Tonry_2003,Blanc_2004,Dahlen_2004,Cappellaro_2005,Scannapieco_2005,Barris_2006,Neill_2006,Neill_2007,Botticella_2007,Poznanski_2007,Dahlen_2008,Kuznetsova_2008,Dilday_2008,Horesh_2008,Graham_2010,Dilday_2010a,Dilday_2010b,Rodney_2010,Barbary_2011a,Barbary_2011b,Graur_2011,Krughoff_2011,Li_2011,Sand_2012,Melinder_2012,Perrett_2012,Graur_2013,Graur_2014,Okumura_2014,Rodney_2014,Graur_2015,Cappellaro_2015,Frohmaier_2019,Perley_2020,Sharon_2022,Desai_2023}). This set represents, to the best of our knowledge, all existing data available for this quantity. All rates are derived for $H_0=70\:km\:s^{-1}\:Mpc^{-3}$, the same value adopted in this work.

\subsection{Long gamma-ray burst rate}\label{sec:lgrb_model}

In order to model the LGRB rate, we first compute the number of LGRB produced per unit star forming mass, $dn_{LGRB}/dM_{\star}$, as:
\begin{equation}\label{eq:stellar_part_LGRB}
\frac{dn_{LGRB}}{dM_{\star}}(Z)=
\begin{cases}
    N_{LGRB}\bigintss_{M_{LGRB}^{low}}^{M_{LGRB}^{up}} dm\:\phi(m) & \text{if}\:\:Z\leq Z_{max} ,\\
    0 & \text{if}\:\:Z>Z_{max}
\end{cases}
\end{equation}
Here, we integrate the IMF over the mass range of LGRB progenitors, and ascribe all uncertainties related to e.g. progenitor rotation, jet-launching conditions and LGRB emission to a normalization factor $N_{LGRB}$ in front of the integral. In addition, we implement a dependence of $dn_{LGRB}/dM_{\star}$ on metallicity, by defining a maximum metallicity $Z_{max}$ above which LGRBs are suppressed. Indeed, the higher the progenitor metallicity, the more mass and angular momentum will be lost via stellar winds. Given that the remnant rotation must be high enough in order to launch the relativistic jet, too high metallicities hinder LGRB emission (e.g. \cite{MacFadyen_1999,Hirschi_2005,Yoon_2005,Woosley_2006b,Yoon_2006})

Since LGRB occurrence depends on the metallicity of the environment where their progenitors form, their study requires a treatment of the galaxy metallicity evolution throughout cosmic history. We resort to a fundamental metallicity relation (FMR), $Z_{FMR}(M_{\star},\psi)$, linking galaxy metallicity to stellar mass and SFR, $\psi$. In particular, we consider the recent determination by \cite{Curti_2020}, taking also into account the offset from the FMR found in \cite{Curti_2023} above $z\sim3$ (we fit their results with a simple power-law function). Due to the intrinsic dependencies of $Z_{FMR}$ on $M_{\star}$ and $\psi$, we build the $Z$-dependent SFRD by convolving GSMFs with a galaxy main sequence (MS), relating stellar mass and SFR, according to the following equation (see also \cite{Boco_2021}):
\begin{equation}\label{eq:Z_SFRD}
\begin{split}
\frac{d^3M_{SFR}}{dtdVd\log Z}(Z,z) = & \int d\log M_{\star}\frac{d^2N}{dVd\log M_{\star}}(M_{\star},z)\\
& \times \int d\log \psi\:\psi \frac{dp}{d\log \psi}(\psi,M_{\star},z)\\
& \times \frac{dp}{d\log Z}(Z,Z_{FMR}(M_{\star},\psi),\sigma_Z)\\
\end{split}
\end{equation}
For coherence with our SMD computation, we consider the GSMFs by \cite{Weaver_2023}, while for the MS we follow \cite{Popesso_2022}. $dp/d\log \psi$ is a log-normal SFR distribution, accounting for both MS and starburst galaxies (SBs). The latter are galaxies with particularly high SFR, which identify a separate region above the MS in the $M_{\star}-\psi$ plane. We treat SBs following \cite{Chruslinska_2021}. $dp/d\log Z(Z,Z_{FMR},\sigma_Z)$ describes a log-normal $Z$-distribution around the FMR, with dispersion $\sigma_Z$.

We then use the result from Equation \ref{eq:Z_SFRD} to compute the fraction of star formation taking place below a certain metallicity at a given redshift, $F$, as:
\begin{equation}
F(z,Z_{max},\sigma_Z)=\dfrac{\bigintss_{-\infty}^{\log Z_{max}} d\log Z \dfrac{d^3M_{SFR}}{dtdVd\log Z}(Z,z)}{\bigintss d\log Z \dfrac{d^3M_{SFR}}{dtdVd\log Z}(Z,z)} 
\end{equation}
$F$ not only depends on redshift and $Z_{max}$, but also on the parameter $\sigma_Z$, i.e. the dispersion in the galaxy metallicity distribution (entering $dp/d\log Z$ in Equation \ref{eq:Z_SFRD}). 

We finally compute the LGRB rate as a function of redshift as:
\begin{equation}\label{eq:LGRB_rate}
R_{LGRB}(z)= \frac{dn_{LGRB}}{dM_{\star}}\times F(z,Z_{max},\sigma_Z)\:\rho_{SFR}(z),
\end{equation}
where $\rho_{SFR}(z)$ is the SFRD from Equation \ref{eq:rho_sfr_uv}.\\

Inferring the LGRB intrinsic rate from observations is more complex than in the case of SNe Ia and CCSNe. Indeed, after the computation of the observed rate from the data, some additional observational biases must be addressed (see e.g. \cite{Chrimes_2019}). On one hand, only LGRBs with jets oriented towards our line of sight can be observed. In order to account for the off-axis population, a correction based on the jet opening angle is required. Moreover, LGRBs are detected up to much higher redshifts than SNe Ia and CCSNe, thus a fraction of the population will inevitably be too faint to be detected with a given instrument. One can overcome this bias by assuming an underlying LGRB luminosity function, and integrating it below the detection threshold of the instrument. Noticeably, \cite{Ghirlanda_2022} derive the intrinsic LGRB rate and luminosity function as a function of redshift by simulating a population of sources with randomly oriented jets, and matching the predictions of their model to several datasets of LGRB observables. These comprise peak flux and energy, fluence, duration, redshift, isotropic equivalent luminosity and energy, and jet opening angle distributions. They also account, for the first time in a LGRB population study, for relativistic effects due to the jet orientation and beaming. We decide to adopt an updated version of their LGRB rate determination, taking into account more recent observations of LGRBs at high redshift (see Appendix \ref{appendix:LGRB_update}), to match to the outcome of our model.

\subsection{MCMC}\label{sec:mcmc_methods}

We perform an MCMC by using the Python package \textit{emcee}\footnote{https://emcee.readthedocs.io/en/stable/} (\cite{Foreman_Mackey_2013}), selecting $2\times 10^2$ walkers. The number of steps needed for the MCMC to converge is approximately $4\times 10^{4}$. The free parameters to be constrained are 13: $\theta$=[$\log\rho_{UV,0}$, $\alpha$, $\beta$, $\gamma$, $\log m_c$, $\xi$, $\log N_{Ia}$, $\log M_{CC}^{up}$, $\sigma_Z$, $\log N_{LGRB}$, $\log Z_{max}$, $\log M_{LGRB}^{low}$, $\log M_{LGRB}^{up}$]. The first 4 parameters define the shape and normalization of $\rho_{UV}(z)$, $N_{Ia}$ is the fraction of stars in the mass range of SN Ia progenitors that actually give rise to a SN Ia, and $M_{CC}^{up}$ is the maximum CCSN progenitor mass; $\log m_c$ and $\xi$ are the IMF characteristic mass and slope; the last 5 parameters are instead related to the metallicity evolution, and the properties of LGRB progenitors: $\sigma_Z$ represents the dispersion of the galaxy metallicity distribution, $N_{LGRB}$ is the fraction of stars with mass in the range of LGRB progenitors, that produce a successful LGRB emission, $Z_{max}$ is the maximum metallicity for LGRBs, and $M_{LGRB}^{low/up}$ set the mass range of LGRB progenitors. We define uniform priors in the ranges $\log\rho_{UV,0}=[25, 26.5]$, $\alpha=[0, 7]$, $\beta=[4, 10]$, $\gamma=[0, 5]$, $\log m_c=[-2,2]$, $\xi=[-5,0]$, $\log N_{Ia}=[-3,0]$, $\log M_{CC}^{up}=[\log(10),\log(150)]$, $\sigma_Z=[0,1]$, $\log N_{LGRB}=[-3,0]$, $\log Z_{max}=[-4,-1]$, $\log M_{LGRB}^{low}=[\log(10),\log(150)]$, $\log M_{LGRB}^{up}=[\log(10),\log(150)]$. For every observable $O$, we then compute the logarithmic likelihood as:
\begin{equation}\label{eq:likelihood}
\ln\mathcal{L}_O(\theta)=-\frac{1}{2} \sum_{i}\left[ \frac{(O(z_i,\theta)-D_i)^2}{s_i^2} + \ln(2\pi s_i^2) \right],
\end{equation}
where the sum runs over the redshifts ${z_i}$ of the corresponding data points. $O(z_i,\theta)$ is the model observable computed at $z_i$, for a given choice of parameters $\theta$, while $D_i$ is the $i$-th data point. In the additional term inside the sum, $s_i^2=\sigma_i^2 + O(z_i,\theta)^2 e^{2\ln f}$, where $\sigma_i$ is the uncertainty associated to $D_i$. $f$ is an additional free parameter that takes into account possible systematics in the data. We allow $\ln f$ to vary uniformly in $[-10,1]$. According to the Bayes theorem, the posterior probability for the choice of parameters $\theta$ can be computed as $\log P(\theta)=\log p(\theta)+\sum_{O} \log\mathcal{L}_O(\theta)$, where $\log p(\theta)$ is 0 if all parameters are within their prior ranges, otherwise it returns $-\infty$. The MCMC takes these quantities as input, and computes the best fit for the model parameters, as well as their marginal and joint posterior probability distributions. We plot the latter quantities via the Python package \textit{GetDist}\footnote{https://getdist.readthedocs.io/en/latest/}, also applying smoothing.\\

As explained in Section \ref{sec:lgrb_model}, due to the complex observational biases linked to LGRB detection, we consider the determination of the LGRB rate as a function of redshift by \cite{Ghirlanda_2022} (more specifically an updated version of their result, see Appendix \ref{appendix:LGRB_update}). For this reason, we compute the likelihood terms related to LGRBs following a conceptually different, but formally identical, approach. In particular, from the updated LGRB rate determination by \cite{Ghirlanda_2022}, we only sample 10 points uniformly spaced in cosmic age. By doing so, we take into account the fact that the number of LGRB observations decreases going to higher redshift, with only a few LGRB detections achieved so far at very-high redshift (e.g. $z>6$). We then define Gaussian priors around each of these points, with errors equal to the $1\sigma$ uncertainty of the LGRB rate determination at the corresponding redshifts. The constraints informed by these priors allow us to estimate the model parameters. We stress that the posterior probability computed in this way is mathematically identical to that shown above, i.e. $\log P(\theta)=\log p(\theta)+\sum_{O} \log\mathcal{L}_O(\theta)$ and Equation \ref{eq:likelihood}, since the additional terms of the Gaussian priors are the same as those in Equation \ref{eq:likelihood}. So the computation of the posterior probability for the parameter space describing LGRBs and the cosmic metallicity evolution is fully consistent with that for the UV luminosity density, local SMD, SN Ia and CCSN rate parameters.

\section{Results}\label{sec:results}

\begin{figure*}
    \includegraphics[width=0.7\textwidth]{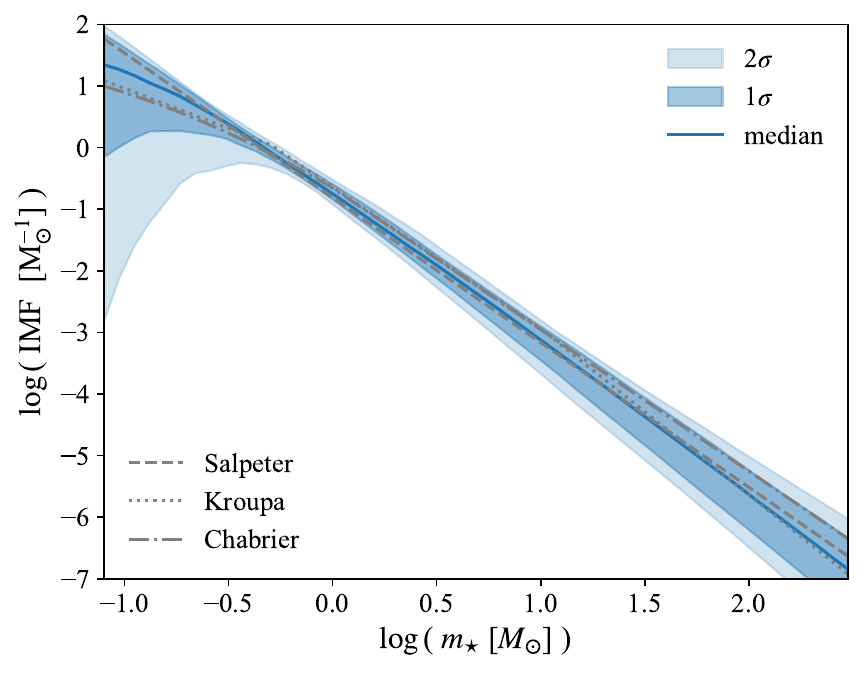}
    \caption{IMF fit as a function of stellar mass. We show the median of the posterior distributions with a blue solid line, as well as the $1$ and $2\sigma$ uncertainty bands around the median. For comparison, we plot the Salpeter, Kroupa, and Chabrier IMFs as grey lines.}
    \label{fig:fit_IMF}
\end{figure*}

\begin{figure*}
    \includegraphics[width=\textwidth]{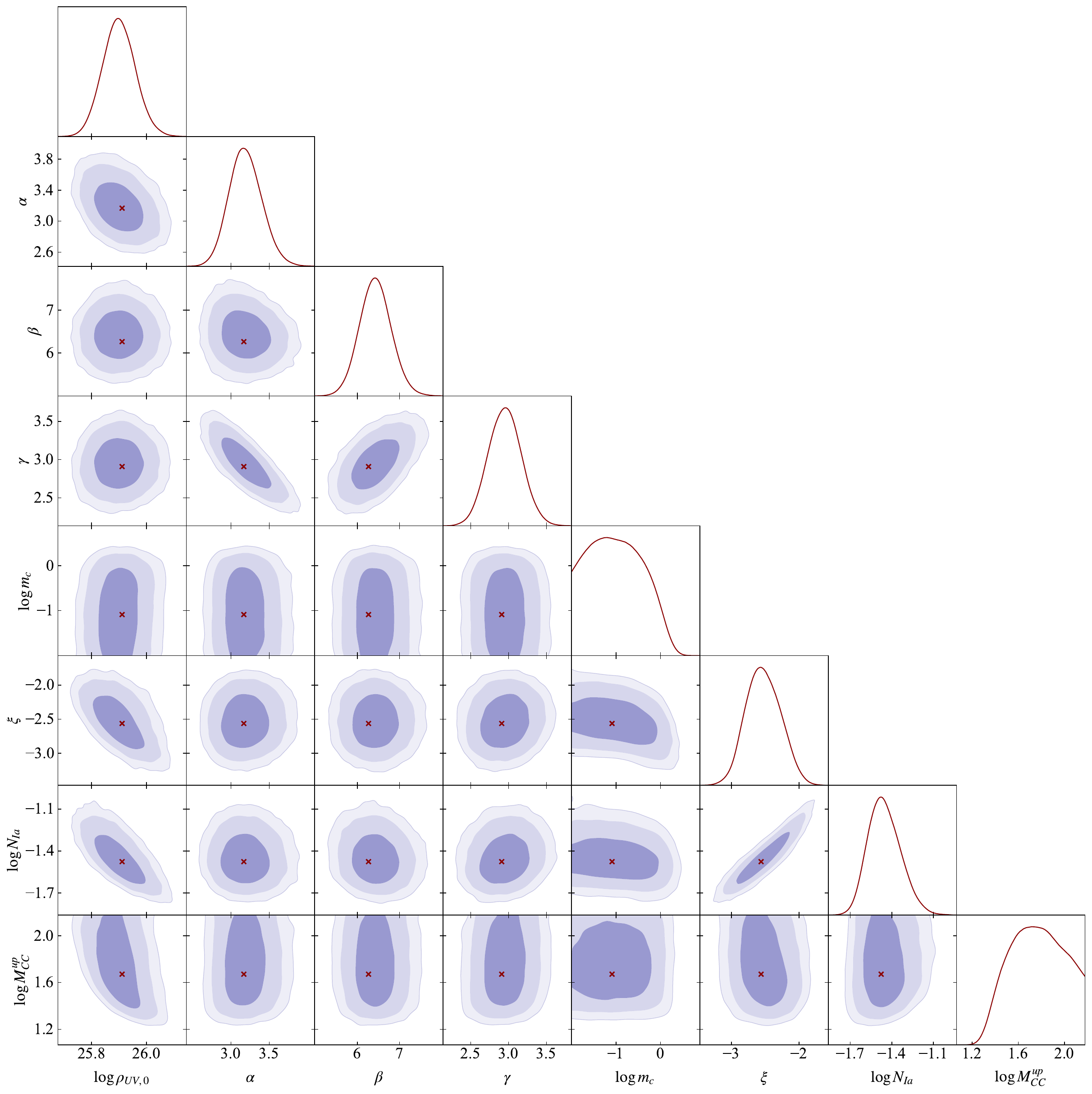}
    \caption{Corner plot showing the individual and joint posterior probability distributions, only for the parameters related to $\rho_{UV}$, $\rho_{\star,0}$, $R_{Ia}$ and $R_{CC}$, including the IMF parameters $\log m_c$ and $\xi$. We show the 1, 2 and $3\sigma$ confidence level regions in progressively lighter shades, respectively. The best-fit values are shown as red crosses.}
    \label{fig:MCMC_ONE}
\end{figure*}

\begin{figure*}
    \centering
    \begin{subfigure}[b]{0.49\textwidth}  
        \centering 
        \includegraphics[width=\textwidth]{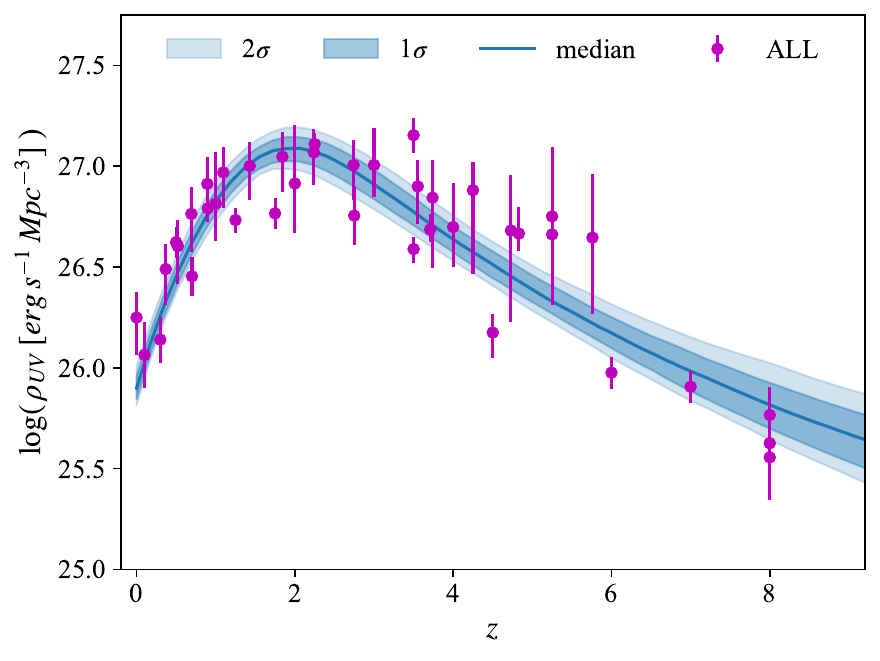}
        \caption[UV luminosity density fit as function of redshift. The data points are from \cite{Gruppioni_2013,Rowan-Robinson_2016,Dunlop_2016,Novak_2017,Liu_2018,Oesch_2018,Bhatawdekar_2019,Gruppioni_2020,Bouwens_2023,Donnan_2023,Harikane_2023}.]%
        {{\small UV luminosity density fit as function of redshift. The data points are from \cite{Gruppioni_2013,Rowan-Robinson_2016,Dunlop_2016,Novak_2017,Liu_2018,Oesch_2018,Bhatawdekar_2019,Gruppioni_2020,Bouwens_2023,Donnan_2023,Harikane_2023}.}}    
    \end{subfigure}
    \hfill
    \begin{subfigure}[b]{0.49\textwidth}
        \centering
        \includegraphics[width=\textwidth]{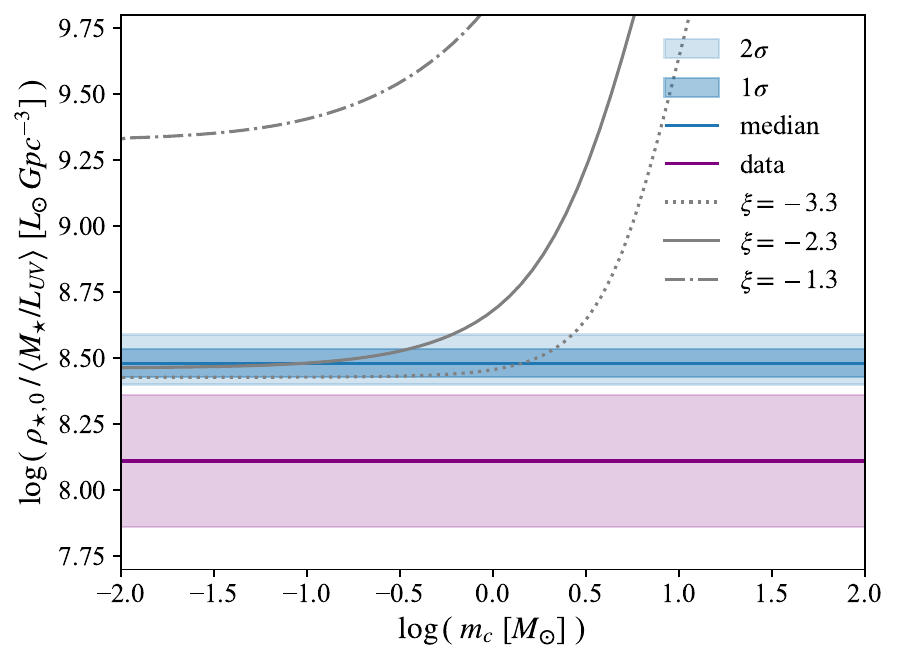}
        \caption[Local SMD fit, compared with the observational determination obtained from \cite{Weaver_2023}.]%
        {{\small Local SMD fit, compared with the observational determination obtained from \cite{Weaver_2023}.}}    
    \end{subfigure}
    \vskip\baselineskip
    \begin{subfigure}[b]{0.49\textwidth}   
        \centering 
        \includegraphics[width=\textwidth]{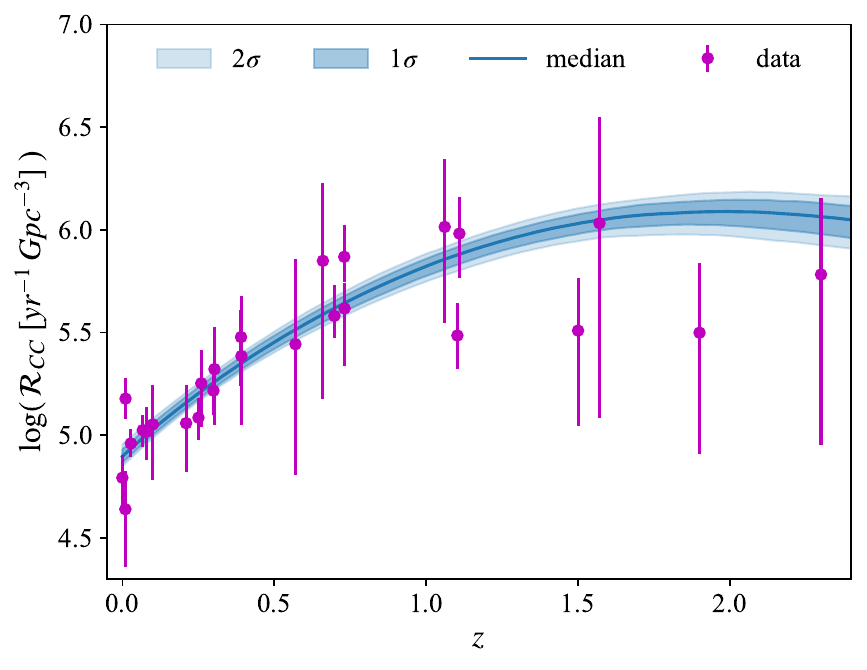}
        \caption[CCSN rate fit as function of redshift. The data points are from \cite{Cappellaro_1999,Cappellaro_2005,Botticella_2007,Bazin_2009,Graur_2011,Li_2011,Dahlen_2012,Mattila_2012,Melinder_2012,Taylor_2014,Cappellaro_2015,Graur_2015,Strolger_2015,Petrushevska_2016,Frohmaier_2020}.]%
        {{\small CCSN rate fit as function of redshift. The data points are from \cite{Cappellaro_1999,Cappellaro_2005,Botticella_2007,Bazin_2009,Graur_2011,Li_2011,Dahlen_2012,Mattila_2012,Melinder_2012,Taylor_2014,Cappellaro_2015,Graur_2015,Strolger_2015,Petrushevska_2016,Frohmaier_2020}.}}    
    \end{subfigure}
    \hfill
    \begin{subfigure}[b]{0.49\textwidth}   
        \centering 
        \includegraphics[width=\textwidth]{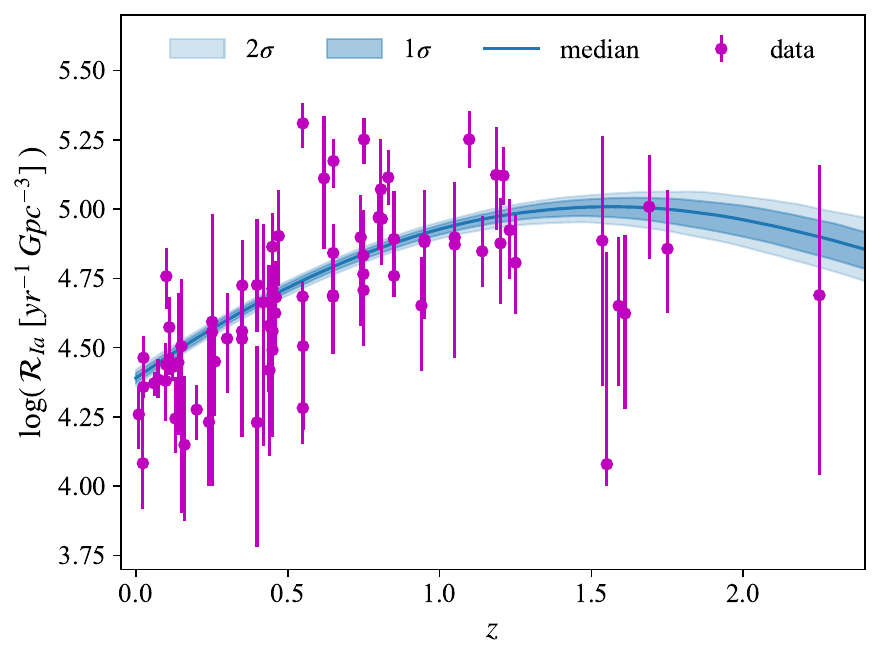}
        \caption[SN Ia rate fit as function of redshift. The data points are from \cite{Pain_1996,Cappellaro_1999,Hardin_2000,Pain_2002,Madgwick_2003,Tonry_2003,Blanc_2004,Dahlen_2004,Cappellaro_2005,Scannapieco_2005,Barris_2006,Neill_2006,Neill_2007,Botticella_2007,Poznanski_2007,Dahlen_2008,Kuznetsova_2008,Dilday_2008,Horesh_2008,Graham_2010,Dilday_2010a,Dilday_2010b,Rodney_2010,Barbary_2011a,Barbary_2011b,Graur_2011,Krughoff_2011,Li_2011,Sand_2012,Melinder_2012,Perrett_2012,Graur_2013,Graur_2014,Okumura_2014,Rodney_2014,Graur_2015,Cappellaro_2015,Frohmaier_2019,Perley_2020,Sharon_2022,Desai_2023}.]%
        {{\small SN Ia rate fit as function of redshift. The data points are from \cite{Pain_1996,Cappellaro_1999,Hardin_2000,Pain_2002,Madgwick_2003,Tonry_2003,Blanc_2004,Dahlen_2004,Cappellaro_2005,Scannapieco_2005,Barris_2006,Neill_2006,Neill_2007,Botticella_2007,Poznanski_2007,Dahlen_2008,Kuznetsova_2008,Dilday_2008,Horesh_2008,Graham_2010,Dilday_2010a,Dilday_2010b,Rodney_2010,Barbary_2011a,Barbary_2011b,Graur_2011,Krughoff_2011,Li_2011,Sand_2012,Melinder_2012,Perrett_2012,Graur_2013,Graur_2014,Okumura_2014,Rodney_2014,Graur_2015,Cappellaro_2015,Frohmaier_2019,Perley_2020,Sharon_2022,Desai_2023}.}}   
    \end{subfigure}
    \caption[Fits obtained for the UV luminosity density, local SMD, CCSN and SN Ia rates as function of redshift, compared with the corresponding observational determinations (purple dots and error bars). The blue line and bands show the median and 1, $2\sigma$ percentiles of the posterior distribution.]
    {\small Fits obtained for the UV luminosity density, local SMD, CCSN and SN Ia rates as function of redshift, compared with the corresponding observational determinations (purple dots and error bars). The blue line and bands show the median and 1, $2\sigma$ percentiles of the posterior distribution.} 
    \label{fig:stellar_transient_fits}
\end{figure*}

In this section, we present the results obtained in this work. Our result on the IMF is displayed in Figure \ref{fig:fit_IMF}. We find the parameters related to LGRBs and the cosmic $Z$ evolution to be somewhat disentangled from the rest of the parameter space. In particular, the IMF estimate is fully determined by the UV luminosity density, local SMD, SN Ia and CCSN rates, and does not require considering also LGRBs. On the other hand, adding the latter transients allows us to also put constraints on the LGRB progenitor properties, and the cosmic $Z$ evolution. For this reason, and for more clarity, we present separately the results obtained for the parameters related to $\rho_{UV}(z)$, $\rho_{\star,0}$, $R_{Ia}(z)$, $R_{CC}(z)$ (Section \ref{sec:IMF_results}, Figure \ref{fig:MCMC_ONE}), and those for $R_{LGRB}(z)$ and the cosmic $Z$ evolution (Section \ref{sec:LGRB_results}, Figure \ref{fig:MCMC_TWO}). For completeness, we also report in Appendix \ref{appendix:more_mcmc} a corner plot with all posteriors found by the MCMC, for the whole parameter space (Figure \ref{fig:MCMC_all}). See Table \ref{tab:MCMC_all} for the median values obtained for all parameters, with $1\sigma$ uncertainties. The total reduced chi square amounts to $\chi^2_{red}=1.17$. In Figures \ref{fig:stellar_transient_fits} and \ref{fig:fit_LGRB_rate}, we show the fits obtained for $\rho_{UV}(z)$, $\rho_{\star,0}$, $R_{CC}(z)$, $R_{Ia}(z)$ and $R_{LGRB}(z)$, in comparison with the corresponding datasets. Finally, in Section \ref{sec:IMF_var} we test the assumption of IMF evolution with $z$.

\subsection{Stellar initial-mass function, type Ia supernovae and core-collapse supernovae}\label{sec:IMF_results}

Our method allows us to obtain for the first time an IMF determination based on the observed rates of stellar transients. We infer an IMF with characteristic mass $\log m_c=-0.99^{+0.52}_{-0.77}$ (i.e. $m_c=0.10^{+0.24}_{-0.08}\:M_{\odot}$), and slope $\xi=-2.53^{+0.24}_{-0.27}$. The IMF slope is remarkably compatible with those typically adopted based on the observed stellar populations in the local Universe. One can see this agreement in Figure \ref{fig:fit_IMF}, where we show our result in comparison with the Salpeter ($\xi=-2.35$, \cite{Salpeter_1955}), Kroupa ($\xi=-2.3,\:-2.7$, \cite{Kroupa_1993,Kroupa_2001,Kroupa_2013}), and Chabrier IMFs ($\xi=-2.3$, \cite{Chabrier_2003}). We note that the uncertainties on $\log m_c$ only enlarge the error band at $\log m_c\lesssim -0.5$, a mass range poorly constrained also by local IMF determinations.

As one can see in Figure \ref{fig:MCMC_ONE}, the $\rho_{UV}$ parameters are particularly well constrained, due to the relatively small uncertainties of the corresponding data, at least at low redshift. The larger scatter and error bars of the SN Ia and CCSN rate data points raises the uncertainties for the other parameters. This combines with the fact that, as described in Section \ref{sec:methods}, their computation requires the conversion from $\rho_{UV}$ to $\rho_{SFR}$, and the inclusion of the IMF. The quantities the MCMC struggles the most to constrain are $\log m_c$ and $\log M_{CC}^{up}$. This is to be expected since we are only able to survey stellar masses down to $3\:M_{\odot}$, i.e. the assumed minimum mass of SN Ia progenitors, while we cannot put any constraint at lower masses. Moreover, the high-mass end of the IMF is only covered by the CCSN and LGRB rate, for which we have a less constraining dataset. Nonetheless, we retrieve estimates also for these quantities, albeit with larger error bars. In particular, we find $\log M_{CC}^{up}=1.76\pm 0.22\:M_{\odot}$, at $1\sigma$ confidence level. This means a range $\sim[35-95]\:M_{\odot}$, and a median value of $\sim 58\:M_{\odot}$. Finally, we obtain an estimate for the fraction of SN Ia progenitors that actually produce a SN Ia, of $\log N_{Ia}=-1.45^{+0.10}_{-0.13}$, or $N_{Ia}=3.5\pm 0.9 \times 10^{-2}$ in non-logarithmic values.

As shown in Figure \ref{fig:stellar_transient_fits}, panel (a), we obtain a $\rho_{UV}$ fit peaking at $z\gtrsim 2$. One can notice how our fit does not intercept some of the data points, that indeed exhibit a significant scatter. This is possibly due to the fact that we combine data from several works, that are independent from each other and are obtained from different galaxy surveys, in different bands. As explained in Section \ref{sec:mcmc_methods}, we account for the possible resulting systematics via an additional parameter $f$ in the computation of the likelihood, the value of which is estimated by the MCMC. The same also applies to our results on the CCSN and SN Ia rates (panels (c) and (d) in Figure \ref{fig:stellar_transient_fits}), that are similarly based on datasets from different works, obtained from independent surveys.

Figure \ref{fig:stellar_transient_fits}, panel (b), shows the fit we obtain for the local SMD. More specifically, the quantity we consider in this plot is the local SMD divided by the mass-to-light ratio, $\langle M_{\star}/L_{UV}\rangle$, from Equation \ref{eq:rho_star_obs}. In this way, we bring all quantities that depend on the IMF parameters on one side of the equation, and leave only the observational constraint on the other side (see Section \ref{sec:smd_method}). We indicate the latter with a purple line and error band. We also show $\rho_{\star,0}/\langle M_{\star}/L_{UV}\rangle$ as a function of $\log m_c$ for three fixed IMF slopes, indicated with grey lines. As one can see, our fit appears systematically higher than the data. This is a manifestation of the well-known mismatch between these two SMD determinations, for which an explanation is yet to be found (see e.g. \cite{Madau_2014} and references therein). Nonetheless, this comparison is enough for us to constrain the IMF, and in particular its slope. Indeed, the plot clearly shows how shallow slopes lead to $\rho_{\star,0}/\langle M_{\star}/L_{UV}\rangle$ values which are too high with respect to the empirical constraint, and are thus excluded. Only slopes $\xi<-2$ are allowed. See Section \ref{sec:disc_IMF} for a deeper discussion about this aspect.

Despite the SN Ia and CCSN data points displaying relatively large error bars, we are still able to constrain the rates of these transients to good accuracy thanks to the interlink with $\rho_{UV}$. As shown in Figure \ref{fig:stellar_transient_fits}, panel (c), $R_{CC}$ peaks at $z\gtrsim 2$, resembling $\rho_{UV}$. Indeed, we remind that $R_{CC}$ is simply a re-normalized $\rho_{SFR}$, the multiplication factor being $dn_{CC}/dM_{\star}$ (Equation \ref{eq:CCSN_rate}). On the contrary, as described in Section \ref{sec:methods}, the computation of $R_{Ia}$ involves the adoption of a DTD. As a consequence, $R_{Ia}$ turns out to peak at lower redshift, closer to $z=1.5$ (Figure \ref{fig:stellar_transient_fits}, panel (d)). One can notice how in both cases the error bands become wider at increasing redshifts, where the data are more sparse and uncertain.

Finally, we find relevant degeneracies between parameters $\rho_{UV,0}$, $\xi$ and $\log N_{Ia}$. The correlation between $\xi$ and $\log N_{Ia}$ can be explained with the fact that higher $\xi$s decrease the number of stars in the range on SN Ia progenitors, between 3 and 8 $M_{\odot}$. As a consequence, $N_{Ia}$ must increase in order to meet the observational constraints on $R_{Ia}$. Moreover, increasing $\rho_{UV,0}$ makes it necessary to decrease $\log N_{Ia}$, and therefore also $\xi$, to avoid producing too many SN Ia events.

It is important to stress that $\rho_{UV}(z)$, $\rho_{\star,0}$, $R_{Ia}(z)$ and $R_{CC}(z)$ are sufficient to provide a robust IMF determination, without the need to resort to LGRBs and the metallicity evolution formalism. Indeed, by running the MCMC without the set of constraints on $R_{LGRB}(z)$, we obtain an IMF compatible with that presented in this section, with parameters $\log m_c=-0.95^{+0.39}_{-0.98}$ and $\xi=-2.71^{+0.30}_{-0.33}$. We report the posterior distributions obtained in this case in Appendix \ref{appendix:more_mcmc} (Figure \ref{fig:MCMC_noLGRB}), as well as the parameter estimates (Table \ref{tab:fit_noLGRB}), for completeness. The reduced chi square is $\chi^2_{red}\sim 1.25$, similar to the case with LGRBs.

\subsection{Long gamma-ray bursts and cosmic metallicity evolution}\label{sec:LGRB_results}

\begin{figure*}
    \includegraphics[width=\textwidth]{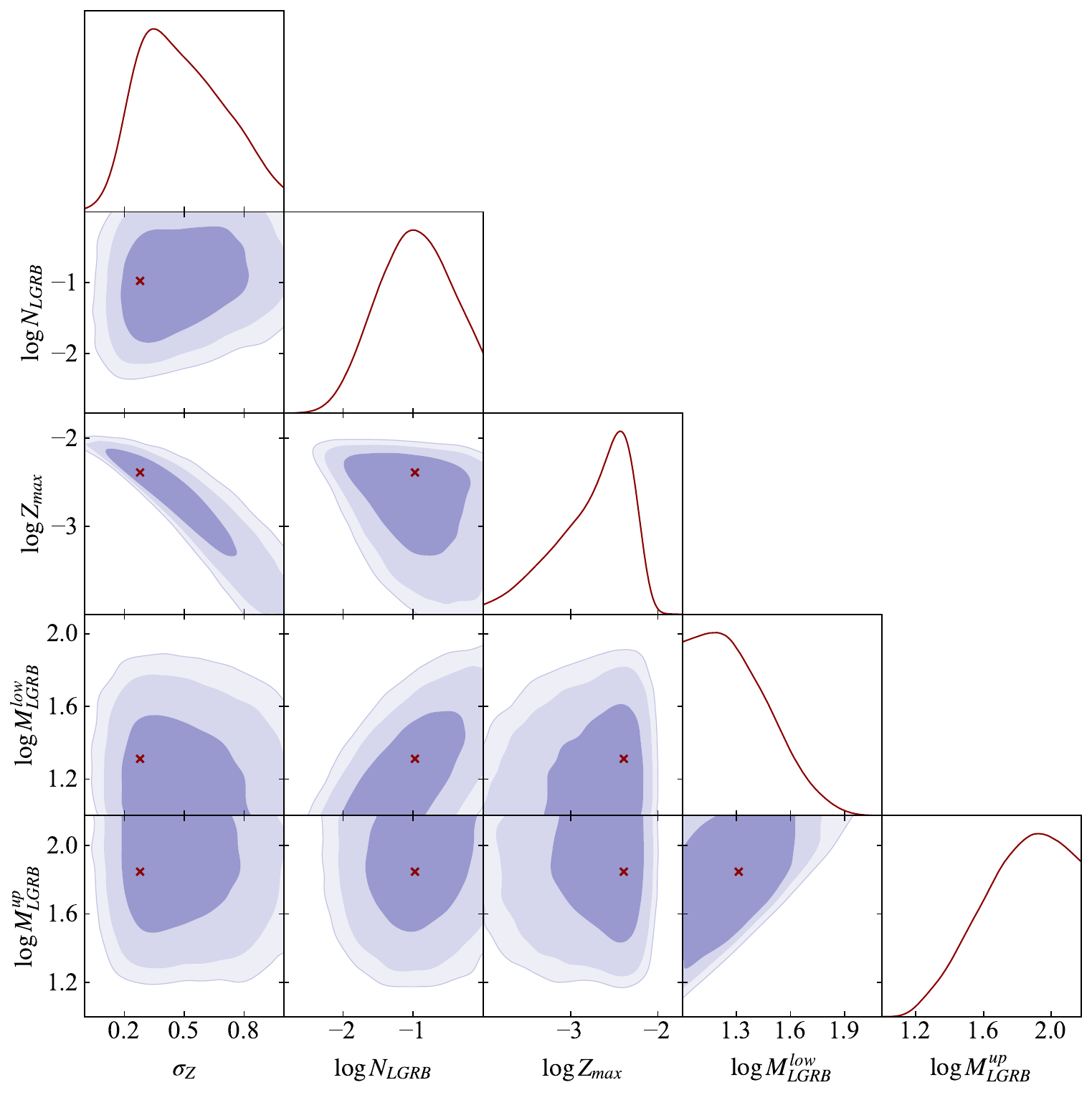}
    \caption{Same as Figure \ref{fig:MCMC_ONE}, only for the parameters related to $R_{LGRB}$ and the cosmic $Z$ evolution.}
    \label{fig:MCMC_TWO}
\end{figure*}

In Figure \ref{fig:MCMC_TWO}, we show the MCMC results for the parameter space describing LGRB progenitors and the cosmic $Z$ evolution. Noticeably, we find an interplay between the maximum metallicity of LGRB progenitors, $Z_{max}$, and the dispersion of the galaxy $Z$ distribution, $\sigma_Z$. In particular, the smaller $\sigma_Z$ is, the higher $Z_{max}$ must be in order to provide enough LGRB rate to match the observations. We are able to constrain $\sigma_Z=0.49^{+0.17}_{-0.27}$, and $\log Z_{max}=-2.75^{+0.54}_{-0.24}$. In non-logarithmic values, $Z_{max}=1.8^{+4.4}_{-0.8}\times 10^{-3}$. Moreover, we find $\log N_{LGRB}=-0.99\pm 0.51$, meaning that a median $\sim 10\%$, and from $\sim 3\%$ to 33$\%$ at $1\sigma$ level, of stars in the mass range of LGRB progenitors satisfy the conditions for LGRB emission. Regarding the mass range of LGRB progenitors, the MCMC is only able to retrieve $\log M_{LGRB}^{low}<1.38$ and $\log M_{LGRB}^{up}>1.71$ (at $1\sigma$), corresponding to $\sim$ $M_{LGRB}^{low}<24\:M_{\odot}$, and $M_{LGRB}^{up}>51\:M_{\odot}$.

\begin{figure*}
    \includegraphics[width=0.7\textwidth]{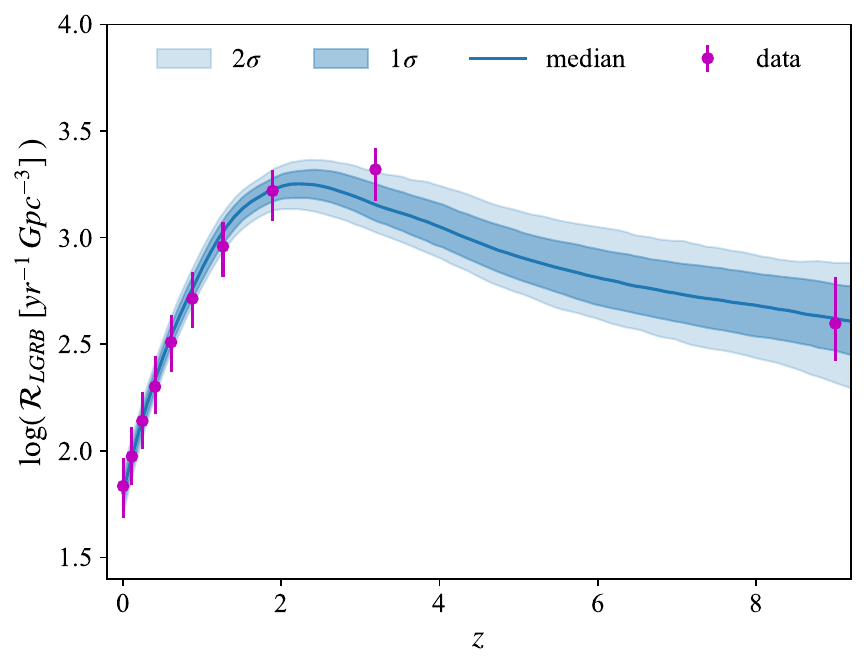}
    \caption{LGRB rate fit as a function of redshift, compared to the constraints from \cite{Ghirlanda_2022}, updated as described in the Appendix (purple dots and error bars). We show the median and 1, $2\sigma$ percentiles with a blue line and bands.}
    \label{fig:fit_LGRB_rate}
\end{figure*}

As one can see from Figure \ref{fig:fit_LGRB_rate}, the $1\sigma$ uncertainty on the LGRB rate fit is somewhat larger than for the quantities showed above, reflecting the uncertainties in the parameter estimates. This is mainly due to the dearth of observational constraints at high redshifts. The LGRB rate peaks between $z=2$ and 3, higher than the $\rho_{UV}$ peak (Figure \ref{fig:stellar_transient_fits}, panel (a)). Indeed, according to our model, $Z_{max}$ completely cuts the $Z$-dependent SFRD distribution above its value, with the effect of shifting the position of the SFRD peak to higher redshifts. See also \cite{Gabrielli_2024}, where this effect is discussed in depth for the cosmic rate of pair-instability supernovae. 

As already pointed out, we find $Z_{max}$ to anti-correlate with $\sigma_Z$. Moreover, $\log N_{LGRB}$ correlates with $\log M_{LGRB}^{low}$, since increasing the lower limit on the mass of LGRB progenitors reduces the value of the integral in Equation \ref{eq:LGRB_rate}, requiring a higher $\log N_{LGRB}$.

\begin{table}
\centering
\captionof{table}{Estimates of the model parameters obtained by combining the observational determinations for $\rho_{UV}(z)$, $\rho_{\star,0}$, $R_{CC}(z)$, $R_{Ia}(z)$ and $R_{LGRB}(z)$. The errors represent the $1\sigma$ uncertainties. We also report the corresponding units in the third column, where dashes indicate dimensionless quantities.}
\begin{tabular} { l c c }
\hline
{\boldmath$\log{\rho_{UV,0}}$} & $25.90\pm 0.06           $ & $\log(erg\:s^{-1}\:Mpc^{-3})$\\
\vspace{0.5mm}
{\boldmath$\alpha         $} & $3.19^{+0.20}_{-0.23}      $ & - \\
\vspace{0.5mm}
{\boldmath$\beta          $} & $6.43\pm 0.38              $ & - \\
\vspace{0.5mm}
{\boldmath$\gamma         $} & $2.95\pm 0.22              $ & - \\
\vspace{0.5mm}
{\boldmath$\log{m_c}      $} & $-0.99^{+0.52}_{-0.77}     $ & $\log(M_{\odot})$\\
\vspace{0.5mm}
{\boldmath$\xi            $} & $-2.53^{+0.24}_{-0.27}     $ & - \\
\vspace{0.5mm}
{\boldmath$\log{N_{Ia}}   $} & $-1.45^{+0.10}_{-0.13}     $ & - \\
\vspace{0.5mm}
{\boldmath$\log{M_{CC}^{up}}$} & $1.76\pm 0.22              $ & $\log(M_{\odot})$\\
\vspace{0.5mm}
{\boldmath$\sigma_Z       $} & $0.49^{+0.17}_{-0.27}      $ & - \\
\vspace{0.5mm}
{\boldmath$\log{N_{LGRB}} $} & $-0.99\pm 0.51             $ & - \\
\vspace{0.5mm}
{\boldmath$\log{Z_{max}}  $} & $-2.75^{+0.54}_{-0.24}     $ & - \\
\vspace{0.5mm}
{\boldmath$\log{M_{LGRB}^{low}}$} & $< 1.38                    $ & $\log(M_{\odot})$\\
\vspace{0.5mm}
{\boldmath$\log{M_{LGRB}^{up}}$} & $> 1.71                    $ & $\log(M_{\odot})$\\
\vspace{0.5mm}
{\boldmath$\ln{f}         $} & $-4.39^{+0.22}_{-0.18}     $ & - \\
\hline
\label{tab:MCMC_all}
\end{tabular}
\end{table}

\subsection{IMF evolution with redshift}\label{sec:IMF_var}

As explained above, in this work we make the assumption of a universal IMF, independent e.g. on redshift, metallicity or SFR. In this section, we suspend this assumption, and explore whether the observational constraints employed in this work can provide any indication of an IMF evolution with redshift. 

We prescribe a redshift dependence for both the IMF characteristic mass and slope. For the former, we adopt a simple power-law dependence, following \cite{Dave_2008}:
\begin{equation}
m_c(z)=m_{c,0}\:(1+z)^{\gamma},
\end{equation}
where $m_{c,0}$ is the IMF characteristic mass at $z=0$, and $\gamma$ describes the evolution with redshift. We instead parametrize the evolution of the IMF slope as in \cite{Fryer_2022}:
\begin{equation}
\xi(z)=
\begin{cases}
    \xi_0 & \text{if}\:\:z<z_0 \\
    \xi_0-f_{imf}(z-z_0) & \text{if}\:\:z\geq z_0 ,
\end{cases}
\end{equation}
where $\xi_0$ is the IMF slope at $z=0$, and $z_0$ is the redshift at which the slope begins to vary. We implement this redshift dependence into the computation of all quantities presented in Section \ref{sec:methods}, leading to $\rho_{UV}(z)$, $\rho_{\star,0}$, $R_{Ia}(z)$, $R_{CC}(z)$ and $R_{LGRB}(z)$. Finally, we run the MCMC with the additional free parameters $[m_{c,0},\:\xi_0,\:\gamma_{imf},\:f_{imf},\:z_0]$, where we substitute $m_c$ from the previous sections with $m_{c,0}$, and $\xi$ with $\xi_0$. Thus, we end up with a total of 16 free parameters, plus the factor $\ln f$ from Section \ref{sec:mcmc_methods}. We note right away that we also tried to run the MCMC with $m_c(z)$ variable and $\xi$ fixed, and with $m_c$ fixed and $\xi(z)$ variable, but we obtained similar results to the case with both parameters variable. Thus we only report the results for the latter case.

\begin{figure*}
    \includegraphics[width=0.7\textwidth]{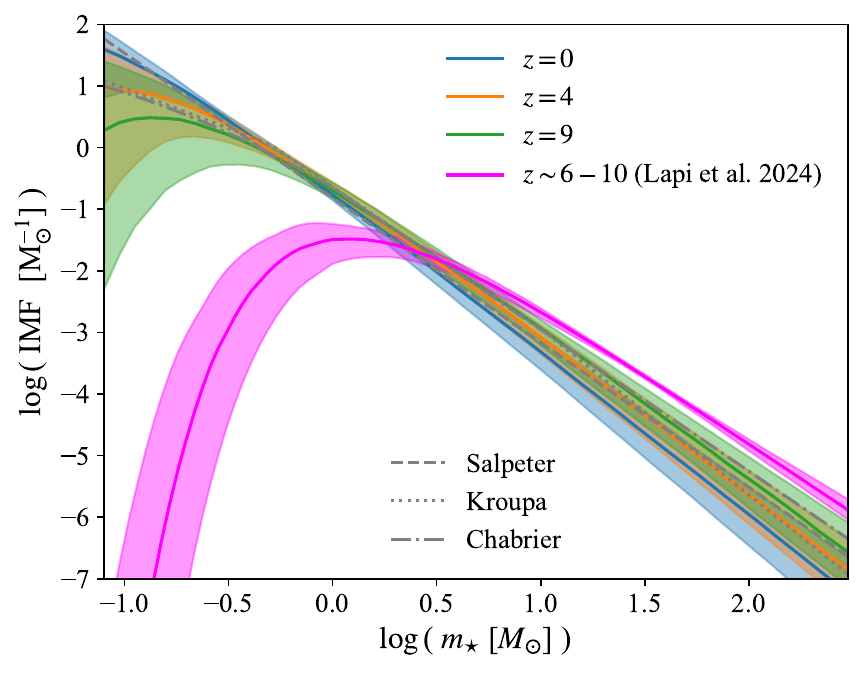}
    \caption{IMF fit as a function of stellar mass, obtained in the case of IMF variable with redshift. In order to show the IMF evolution, we plot the results obtained for $z=0,\:z=4$ and $z=9$ (blue, orange and green lines respectively). Solid lines indicate the median of the posterior distributions, while the shaded areas correspond to the $1\sigma$ errors. For comparison, we plot the Salpeter, Kroupa and Chabrier IMFs as grey lines. We also show the IMF at the epoch of reionization ($z\sim 6-10$) derived in \cite{Lapi_2024}, assuming the \cite{Mitra_2023} escape fraction of ionizing photons. See the text for a comparison with our results.}
    \label{fig:fit_IMF_IMFvar}
\end{figure*}

\begin{figure*}
    \includegraphics[width=\textwidth]{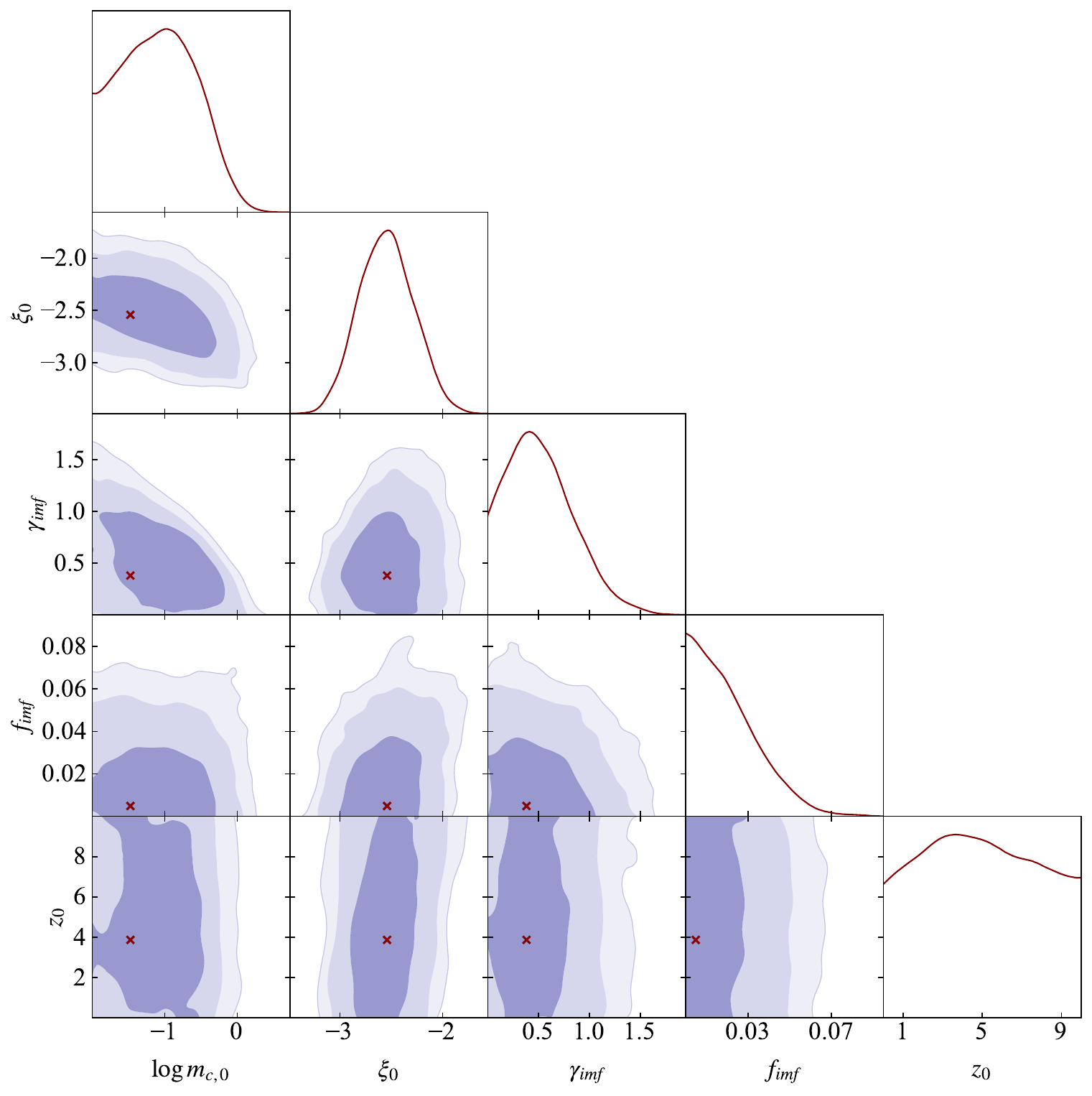}
    \caption{Corner plot with individual and joint posterior probabilities obtained in the case of IMF variable with redshift, only for the subset of the parameter space describing the IMF and its evolution. For other details see the caption of Figure \ref{fig:MCMC_ONE}.}
    \label{fig:MCMC_IMFvar}
\end{figure*}

Figure \ref{fig:fit_IMF_IMFvar} shows the IMF fit we obtain at $z=0$, $z=4$ and $z=9$, in order to appreciate the IMF dependence on redshift. In Figure \ref{fig:MCMC_IMFvar} we show the posterior probabilities we obtain for the parameter space related to the IMF evolution, while in Table \ref{tab:MCMC_IMFvar} we report the corresponding parameter estimates. For the complete corner plot over the whole parameter space, and the table with all parameter estimates, see Appendix \ref{appendix:more_results_IMFvar}. The reduced chi square is $\chi^2_{red}\sim 1.25$, similar to that obtained in the case of universal IMF.

According to our results, the IMF slope appears to be non-evolving with redshift, with an estimate of $f_{imf}<0.026$ at $1\sigma$ confidence level. The parameter $z_0$ instead remains unconstrained. On the other hand, the estimate we find for $\gamma_{imf}=0.53^{+0.21}_{-0.43}$ indicates a mild evolution of the IMF characteristic mass with redshift. Both slope and characteristic mass estimates at $z=0$, $\xi_0=-2.55^{+0.25}_{-0.29}$ and $\log m_{c,0}=-1.08\pm 0.51$, are compatible with those obtained assuming a universal IMF. The same is valid also for the other parameter estimates, see Figure \ref{fig:MCMC_all_IMFvar} and Table \ref{tab:MCMC_IMFvar} in Appendix \ref{appendix:more_results_IMFvar}. 

In Figure \ref{fig:fit_IMF_IMFvar}, we also show the result for the IMF at the epoch of reionization, i.e. $z\sim 6-10$, from \cite{Lapi_2024}, obtained by combining constraints from astrophysical and cosmological data. In particular, we report their result for the \cite{Mitra_2023} escape fraction of ionizing photons. Our IMF slopes are compatible, and we agree in finding no evidence of evolution for this parameter. On the other hand, the \cite{Lapi_2024} IMF exhibits a significantly higher characteristic mass than what found here. We stress that the set of observational constraints adopted in this work are not enough to make an accurate assessment regarding the IMF evolution with redshift. In particular, we are not able to sample low masses relevant to constrain $m_c$, especially at high redshifts where we only have LGRB data. On the other hand, \cite{Lapi_2024} focus on the reionization redshifts $z\sim 6-10$, where they impose more robust and stringent constraints. Therefore, this comparison is not meant as rigorous, and should be taken with caution.

Despite the uncertainties highlighted above, this exploration shows that it might be possible in the future to use this approach to further study the IMF evolution throughout cosmic history, and/or galactic environment, thanks to more and more constraining data.

\begin{table}
\centering
\captionof{table}{Estimates of the parameters describing the IMF evolution, obtained by implementing an IMF dependence on redshift. The errors represent the $1\sigma$ uncertainties. The third column reports the corresponding units, where dashes indicate dimensionless quantities. Note how the parameter $z_0$ remains unconstrained by the MCMC.}
\begin{tabular} { l c c }
\hline
{\boldmath$\log{m_{c,0}}  $} & $-1.08\pm 0.51             $ & $\log(M_{\odot})$\\
\vspace{0.5mm}
{\boldmath$\xi_0          $} & $-2.55^{+0.25}_{-0.29}     $ & - \\
\vspace{0.5mm}
{\boldmath$\gamma_{imf}   $} & $0.53^{+0.21}_{-0.43}      $ & - \\
\vspace{0.5mm}
{\boldmath$f_{imf}        $} & $< 0.026                  $ & - \\
\vspace{0.5mm}
{\boldmath$z_0            $} & -                         & - \\
\hline
\label{tab:MCMC_IMFvar}
\end{tabular}
\end{table}

\section{Discussion}\label{sec:discussion}

\subsection{Initial-mass function}\label{sec:disc_IMF}

In this work we find that the observational constraints coming from the UV luminosity density, the local SMD, the SN Ia and CCSN rates as a function of redshift, allow us to obtain a new determination of the stellar IMF. Remarkably, our result is compatible with the most common IMFs traditionally assumed, inferred from stellar observations in the local Universe, e.g. the Salpeter, Kroupa, Chabrier IMFs. This might indicate that these IMFs describe stellar populations not only locally, but at least up to Cosmic Noon (i.e. $z\sim 2-3$), up to which we have SN Ia and CCSN rate data. Since we do not make any assumption on the IMF, outside of a flexible Larson-shaped parametrization, our result is directly informed by the observed stellar transient rates.\\

We find the local SMD to play an important role in constraining the IMF, as can be appreciated by looking at Figure \ref{fig:stellar_transient_fits}, panel (b). As one can see, shallow slopes are excluded since they produce too high $\rho_{\star,0}/\langle M_{\star}/L_{UV}\rangle$. This limits to slopes $\xi\lesssim -2$. If we remove the constraint on the local SMD, a second peak appears in the $\xi$ posterior distribution. The original peak remains around $\xi\sim -2.5$, while the second one lies around $\xi\sim -1.5$. We believe the latter to be a non-physical solution allowed by the data. Including the local SMD in our dataset completely eliminates this solution. Indeed, while shallower IMF slopes increase $k_{UV}$, leading to lower SFRD (Equation \ref{eq:rho_sfr_uv}) and lower local SMD (Equation \ref{eq:rho_star}), the mass-to-light ratio $\langle M_{\star}/L_{UV}\rangle$ decreases more steeply for $\xi\gtrsim -2$. As a consequence, $\rho_{\star,0}/\langle M_{\star}/L_{UV}\rangle$ starts to rapidly increase at $\xi\gtrsim -2$, becoming too high with respect to the observational constraint (Equation \ref{eq:rho_star_obs}, where we moved $\langle M_{\star}/L_{UV}\rangle$ to the other side of the equation). The result is that IMFs with slope below $\sim -2$ are ruled out.\\

By adopting a constant IMF, we are assuming that the IMF is universal throughout the redshift range considered in this work. This is a common approximation, despite a number of works exploring the IMF dependence on the environment, e.g. redshift, metallicity and SFR (e.g. \cite{Strolger_2015,Stanway_2019,Ziegler_2022,Li_2023,Martin-Navarro_2023,Maksymowicz-Maciata_2024,Wainer_2024}). The fact that we obtain an IMF compatible with local stellar observations might support the IMF universality. In order to further explore this point, we try to implement a redshift dependence for the IMF, and see if the MCMC can find any evidence of evolution. We find that the observables employed here do not support any significant variation of the IMF slope with redshift. On the other hand, they indicate a mild increase of the IMF characteristic mass with redshift. The uncertainties of the datasets we employ, and the dearth of constraints at high redshift, prevent us from saying anything more about the IMF evolution. Nonetheless, our results show how this treatment could enable to obtain more stringent constraints in the future, when many more observations will be available.\\

\subsection{Type Ia and core-collapse supernova progenitors}

As described in Section \ref{sec:methods}, in order to be able to constrain the IMF and the other parameters of interest, we decide to agnostically account for all uncertainties related to the binary nature of SNe Ia with a constant factor $N_{Ia}$ in front of the IMF integral in Equation \ref{eq:SNIa_rate}. $N_{Ia}$ represents the fraction of stars with progenitor mass in the range for SN Ia, which actually explode as SN Ia. It also agnostically accounts for the fraction of stars lying in binaries, $f_{bin}$. We find $N_{Ia}=3.5\pm 0.9 \times 10^{-2}$, meaning that roughly 2 to $5\%$ of stars give rise to a SN Ia, given that their mass is in the SN Ia progenitor range. As already shown in Section \ref{sec:results}, $\log N_{Ia}$ is degenerate with the IMF slope, motivating the uncertainties on its estimate.\\

If we multiply $N_{Ia}$ by the IMF integral over the mass range of SN Ia progenitors, $\int_{3\:M_{\odot}}^{8\:M_{\odot}}\phi(m)\:dm \sim 0.016\:M_{\odot}^{-1}$ (for our median $\log m_c$ and $\xi$), we get the number of SN Ia produced per unit stellar mass, with respect to the entire stellar population, i.e. $dn_{Ia}/dM_{\star}$. Using our $N_{Ia}$ estimate, we get $dn_{Ia}/dM_{\star}=5.6\pm 1.4 \times 10^{-4}\:M_{\odot}^{-1}$. This is in accordance with previous works that compute this quantity using approaches similar to ours, where $dn_{Ia}/dM_{\star}$ generally falls around $1\times 10^{-3}\:M_{\odot}^{-1}$ (\cite{Madau_1998,Dahlen_2004,Botticella_2007,Valiante_2009,Horiuchi_2010,Cappellaro_2015,Petrecca_2024}). The lower and higher values are, respectively, $3.4\times 10^{-4}\:M_{\odot}^{-1}$ (\cite{Botticella_2007}), and $1.14\times 10^{-2}\:M_{\odot}^{-1}$ (\cite{Valiante_2009}), accounting for uncertainties. The most distant results from $1\times 10^{-3}\:M_{\odot}^{-1}$ are due to the adoption of extreme DTDs and SFRDs. In comparing our results with the literature, when needed we convert their results to the quantity $dn_{Ia}/dM_{\star}$ accounting for the IMF adopted in each work.\\

$f_{bin}$ is still quite uncertain (e.g. \cite{Sana_2012}). \cite{Moe_2017} and \cite{Winters_2019} measure the local $f_{bin}$ as a function of stellar mass, via direct observations in the Solar neighborhood. According to their results, stars in the mass range of SN Ia progenitors (i.e. $3-8\:M_{\odot}$ as assumed in this work) would have a probability between roughly 0.5 and 0.7 of lying in a binary (increasing with mass). Assuming that this measured $f_{bin}$ behaviour is valid also at higher redshifts, we fix $f_{bin}$ and compute the quantity $b_{Ia}=N_{Ia}/f_{bin}$. $b_{Ia}$ describes the fraction of stars in the mass range of SN Ia progenitors, which produce SN Ia, given that they are in a binary system. If we fix $f_{bin}=0.5$, we get $b_{Ia}=7.0\pm 1.8\times 10^{-2}$. Fixing instead $f_{bin}=0.7$ gives $b_{Ia}=5.0\pm 1.3 \times 10^{-2}$. This means that, accounting for the uncertainties on our $N_{Ia}$ estimate and on $f_{bin}$, roughly from $4\%$ to $10\%$ of stars lying in binaries, with mass in the SN Ia progenitor range, actually give rise to a SN Ia, due to the processes happening in the binary.\\

As our results suggest, this kind of studies offers the opportunity to constrain the properties of SN Ia progenitors, and delve into aspects such as the relative contribution to the SN Ia rate coming from the SD and DD scenarios (see also e.g. \cite{Palicio_2024}), and the evolution of WD binary systems.\\

The upper mass limit of CCSN progenitors, $M_{CC}^{up}$, is very uncertain. Indeed, it is still not clear what is the transition mass between type II and type Ib/c CCSN subtypes, and at what masses failed SNe start to take place. $M_{CC}^{up}$ might lie between $\sim$20-25 $M_{\odot}$, all the way up to 100 $M_{\odot}$. Our results point to $M_{CC}^{up}\in[35-95]\:M_{\odot}$, with a median value of $\sim 58\:M_{\odot}$. Even though the observational constraints we employ do not allow to add much to what already known, this shows the potential of this type of studies for estimating this quantity, especially with the growth of observational data in the future (e.g. \cite{Refregier_2010,Laureijs_2011,Feindt_2019,Regos_2019,Rose_2021}). It is worth mentioning that the lack of CCSN observations from progenitors above $\sim 20-25\:M_{\odot}$, might be due to observational biases. In particular, CCSNe from more massive progenitors might be too dim to be detected (e.g. \cite{Smartt_2015,Coughlin_2018,Antoni_2023}). In this case, it would be more correct to consider our parameter $M_{CC}^{up}$ as the maximum mass of CCSN progenitors \textit{that can be observed}. A possible addition to this work would thus be to correct for this observational bias, and find the maximum mass of all CCSN progenitors, detectable or undetectable.\\

\subsection{Long gamma-ray bursts and cosmic metallicity evolution}

We see that adding the LGRB rate, and prescribing a cosmic $Z$-evolution recipe, allows to put constraints on relevant quantities related to LGRB progenitors and the cosmic metallicity evolution. LGRBs are expected to arise only at metallicities lower than a certain value $Z_{max}$. Remarkably, we are able to constrain this quantity to $Z_{max}=1.8^{+4.4}_{-0.8}\times 10^{-3}$, or $12+\log$(O/H)$=7.83^{+0.54}_{-0.24}$. As explained in Section \ref{sec:methods}, we find an interplay between $Z_{max}$ and the dispersion in the galaxy metallicity distribution, $\sigma_Z$, and we are able to constrain the latter to $\sigma_Z=0.49^{+0.17}_{-0.27}$. \cite{Ghirlanda_2022} find a threshold value of $12+\log$(O/H)=8.6, or $Z_{max}\sim 1\times 10^{-2}$, that is somewhat higher than our result. Also LGRB host galaxy studies tend to show a preference for $Z_{max}$ from $\sim 6\times 10^{-3}$ up to solar values, $\sim 2\times 10^{-2}$ (e.g. \cite{Kocevski_2009,Vergani_2015,Perley_2016,Vergani_2017,Palmerio_2019}). On the other hand, the values generally found by single stellar evolution simulations are more consistent with our results, $Z_{max}\sim 2-6\times 10^{-3}$ (e.g. \cite{Langer_2006,Woosley_2006,Yoon_2006}). Accounting for the possible binary nature of LGRB progenitors in simulations might significantly change the latter results. E.g., rotating stars in binaries can keep most of their initial surface velocity, or even increase it, and might thus give rise to LGRBs even at higher metallicities (see e.g. \cite{Palmerio_2019} and references therein). Moreover, the observed LGRB dependence on metallicity might also be due to an underlying IMF variability, rather than simply to the progenitor evolution (e.g. \cite{Palmerio_2019,Fryer_2022}).\\

Similarly as discussed for the SN Ia rate, we account for all uncertainties regarding LGRB progenitors and emission by putting a constant factor $N_{LGRB}$ in front of the integral in Equation \ref{eq:LGRB_rate}. In this way, we agnostically account for all conditions related to the progenitor rotation, the mass of the accretion disk that forms after the progenitor explodes as a SN, the nature of the remnant, the launch of the relativistic jet, and finally the conversion of its energy into LGRB emission. Moreover, we account for the possible binary nature of the progenitor, which might favour LGRB emission. We find $N_{LGRB}=10^{+23}_{-7}\times 10^{-2}$, meaning that, as a median value, $\sim10\%$ of stars in the mass range of LGRB progenitors satisfy the conditions for LGRB emission. Within $1\sigma$ confidence level, this percentage ranges from $\sim 3\%$ to $\sim 33\%$. As one can see from Figure \ref{fig:MCMC_TWO}, the uncertainties on this quantity are mostly due to those on the LGRB progenitor mass range. Indeed, despite the MCMC being able to retrieve a fit for $M_{LGRB}^{low}$ and $M_{LGRB}^{up}$, the uncertainties are significant: $M_{LGRB}^{low}<24\:M_{\odot}$, and $M_{LGRB}^{up}>51\:M_{\odot}$, at $1\sigma$. Given the huge uncertainty on these quantities in the literature, these results still give us an indication on where LGRB progenitor masses might lie.\\

There have been a number of associations of LGRBs with type Ib/c SNe, suggesting a common progenitor. SN Ib/c, or "stripped-envelope" supernovae, are a sub-type of CCSNe arising from "Wolf-Rayet" stars that lost their envelope. As already presented, we find a maximum mass of CCSN progenitors, comprising both type II and Ib/c SNe, of $\sim 58^{+37}_{-23}\:M_{\odot}$ at $1\sigma$. On the other hand, we get estimates of $M_{LGRB}^{low}<24\:M_{\odot}$, and $M_{LGRB}^{up}>51\:M_{\odot}$. Given the overlap of the mass ranges of CCSN and LGRB progenitors, and in particular between $M_{CC}^{up}$ and $M_{LGRB}^{up}$, our results appear to be compatible with LGRB and SN Ib/c sharing a common progenitor, even though the uncertainties we find do not allow us to make any further statement. Nonetheless, this treatment suggests the potential of this type of study to unveil the nature of transient progenitors, any possible association among them, and other uncertain aspects such as failed SNe.\\

Finally, one of the open questions about LGRB progenitors revolves around the nature of the remnant, that is believed to be either a BH or a NS. According to the "collapsar" scenario, a highly-spinning BH remnant surrounded by a sufficiently massive accretion disk would be a necessary condition to launch a relativistic jet. On the other hand, the "magnetar" scenario predicts the remnant to be a highly-magnetized, and rapidly-spinning, NS. In principle, one could use the results obtained in this work to distinguish between these two scenarios, based on the constraints on the LGRB progenitor masses. E.g., if $M_{LGRB}^{low}$ turned out to lie above the expected masses of NS progenitors, than the "magnetar" hypothesis would be excluded in favour of the BH remnant one. However, the high uncertainties we obtain for these quantities prevent us from saying anything about this matter. It would be nonetheless intriguing to explore this aspect once more constraining data will be available in the future (e.g. \cite{White_2021,Fryer_2022,Bozzo_2024}).\\

\subsection{Variation on the cosmic $Z$-evolution prescription}\label{sec:FMR_variations}

\begin{figure*}
    \centering
    \begin{subfigure}[b]{0.49\textwidth}  
        \centering 
        \includegraphics[width=\textwidth]{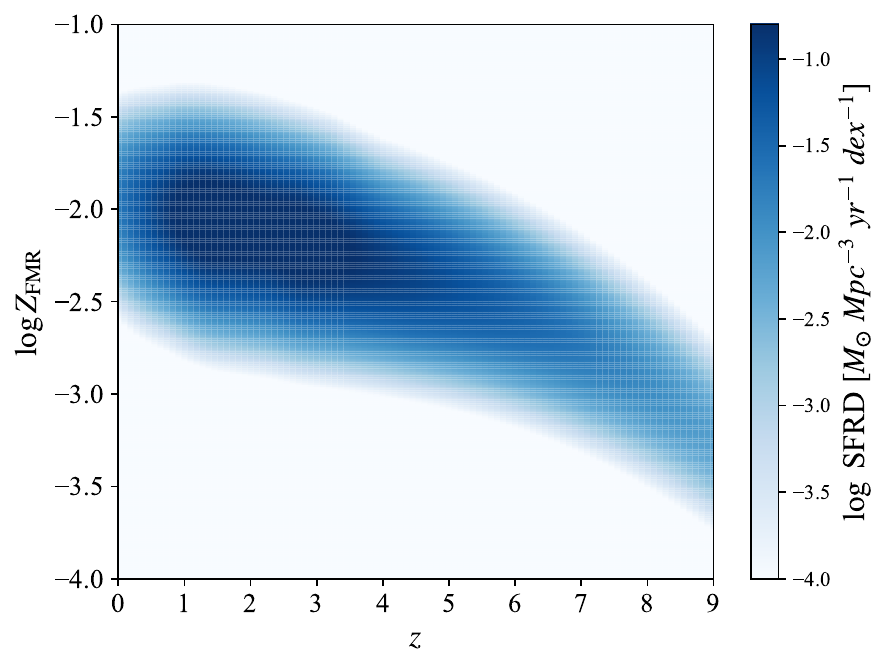}
        \caption[Z-dependent SFRD for the FMR by \cite{Curti_2020,Curti_2023}]%
        {{\small Z-dependent SFRD for the FMR by \cite{Curti_2020,Curti_2023}}}
        \label{fig:FMR_Curti2023}
    \end{subfigure}
    \hfill
    \begin{subfigure}[b]{0.49\textwidth}
        \centering
        \includegraphics[width=\textwidth]{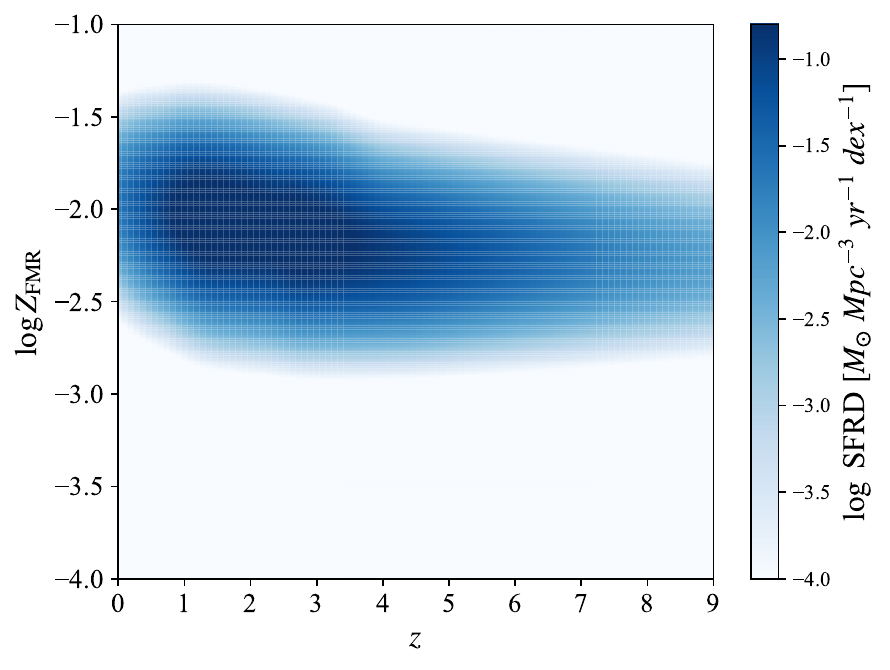}
        \caption[Z-dependent SFRD for the FMR by \cite{Curti_2020}]%
        {{\small Z-dependent SFRD for the FMR by \cite{Curti_2020}}}   
        \label{fig:FMR_Curti20}
    \end{subfigure}
    \vskip\baselineskip
    \begin{subfigure}[b]{0.46\textwidth}  
        \centering 
        \includegraphics[width=\textwidth]{CORR_z9_FINAL_PLOTS_meanerrs_lesspoints_CONVTEST_LGRBrate_fit.pdf}
        \caption[LGRB rate fit for the FMR by \cite{Curti_2020,Curti_2023}]%
        {{\small LGRB rate fit for the FMR by \cite{Curti_2020,Curti_2023}}}
        \label{fig:LGRBrate_Curti2023}
    \end{subfigure}
    \hfill
    \begin{subfigure}[b]{0.46\textwidth}
        \centering
        \includegraphics[width=\textwidth]{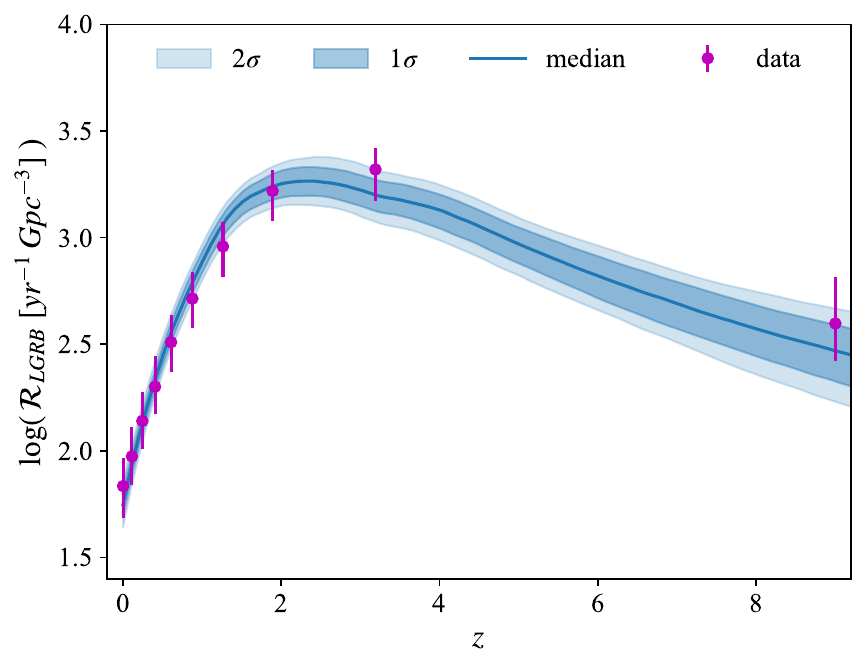}
        \caption[LGRB rate fit for the FMR by \cite{Curti_2020}]%
        {{\small LGRB rate fit for the FMR by \cite{Curti_2020}}}   
        \label{fig:LGRBrate_Curti20}
    \end{subfigure}
    \vskip\baselineskip
    \begin{subfigure}[b]{0.46\textwidth}  
        \centering 
        \includegraphics[width=\textwidth]{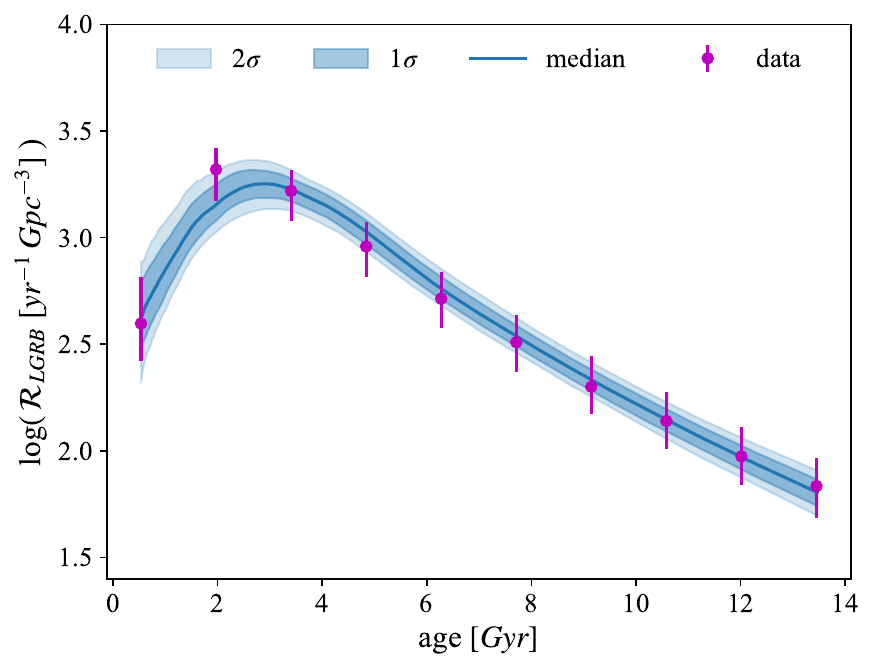}
        \caption[Same as panel (c), as function of cosmic age.]%
        {{\small Same as panel (c), as function of cosmic age.}}    
        \label{fig:LGRBrate_Curti2023_age}
    \end{subfigure}
    \hfill
    \begin{subfigure}[b]{0.46\textwidth}
        \centering
        \includegraphics[width=\textwidth]{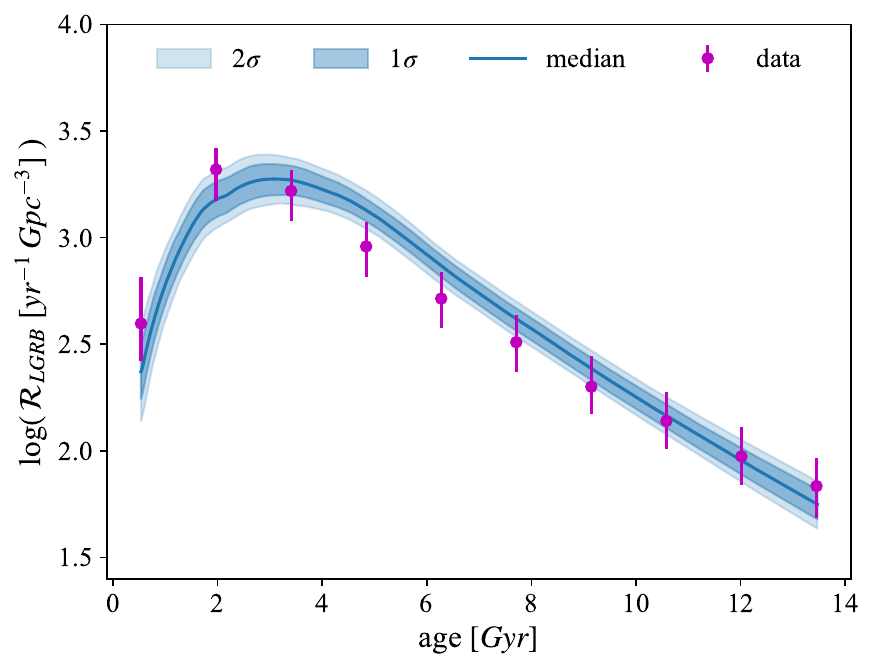}
        \caption[Same as panel (d), as function of cosmic age.]%
        {{\small Same as panel (d), as function of cosmic age.}}    
        \label{fig:LGRBrate_Curti20_age}
    \end{subfigure}
    \caption[Comparison between the $Z$-dependent SFRDs (upper panels) and LGRB rate fits we obtain by adopting the FMR by \cite{Curti_2020} plus the offset from \cite{Curti_2023} (left panels), and the original FMR by \cite{Curti_2020} without the offset from \cite{Curti_2023} (right panels). We show the LGRB fits both as function of redshift (middle panels) and cosmic age (lower panels). $\sigma_Z$ is fixed to 0.15 in the upper panels.]
    {\small Comparison between the $Z$-dependent SFRDs (upper panels) and LGRB rate fits we obtain by adopting the FMR by \cite{Curti_2020} plus the offset from \cite{Curti_2023} (left panels), and the original FMR by \cite{Curti_2020} without the offset from \cite{Curti_2023} (right panels). We show the LGRB fits both as function of redshift (middle panels) and cosmic age (lower panels). $\sigma_Z$ is fixed to 0.15 in the upper panels.} 
    \label{fig:FMR_variation} 
\end{figure*}

As explained above, in order to compute the LGRB rate as function of redshift, we adopt the FMR by \cite{Curti_2020}, further implementing the offset from the FMR found in \cite{Curti_2023}. In this section, we compare our results to the case where we do not consider the offset from \cite{Curti_2023}, and only adopt the original \cite{Curti_2020} FMR (Figure \ref{fig:FMR_variation}). To ease the reading, we will refer to the two cases as C20 and C20+23. As one can see in Table \ref{tab:FMR_Curti20} in Appendix \ref{appendix:FMR_variations}, the parameter estimates obtained in the C20 case are compatible with those obtained in this work (compare with Table \ref{tab:MCMC_all}).

As shown in panels (a) and (b) of Figure \ref{fig:FMR_variation}, the galaxy metallicity decreases much slower with redshift for the C20 FMR, with respect to the C20+23 one. By looking at panels (c) and (d), we see that both FMRs seem to produce relatively good fits to the observational points. If we compute the corresponding chi squares, we actually find that the C20 case is somewhat disfavoured, with $\chi_{LGRB}^2=3.5$ as opposed to $\chi_{LGRB}^2=1.5$ for the C20+23 FMR. This can be better appreciated by looking at panels (e) and (f), where we show the LGRB fit as function of cosmic age, instead of redshift. Indeed the fit somehow struggles to intercept some of the points in the C20 case (panel (f)). Even though the uncertainties and low significance of these results prevent us from drawing any robust conclusion, they might suggest the potential of this approach to discriminate between different $Z$ evolutions in the future, with more LGRB observations available at high redshift. For example, if the results shown in this section were reproduced at higher significance level, then FMRs with a steeper decrease with redshift would be favoured with respect to more constant ones.

\section{Conclusions}\label{sec:conclusions}

In this work, we seek to find a new determination of the stellar IMF, inferred from the observed rates of stellar transients. We adopt a parametric approach in order to model the UV luminosity density, local SMD, SN Ia, CCSN and LGRB rates as function of redshift, and constrain the model parameters in order to reproduce the observational data available for these quantities. Remarkably, we obtain an IMF compatible with those typically assumed based on stellar observations in the local Universe, e.g. Salpeter, Kroupa, Chabrier. In particular, we infer an IMF slope of $\xi=-2.53^{+0.24}_{-0.27}$, consistent with the usual values of -2.3 (Chabrier, Kroupa), -2.35 (Salpeter), and -2.7 (Kroupa, Scalo). We stress that our determination is completely independent from all previous ones. Moreover, our parametric approach allows us to keep the number of physical assumptions at the minimum, leading to a fully observation-driven IMF determination. We also find that, according to our results, a fraction $N_{Ia}=3.5\pm 0.9 \times 10^{-2}$ of stars in the mass range of SN Ia progenitors actually explode as SN Ia, due to the processes happening in the binaries.

The LGRB rate is not crucial to infer the IMF, which can be robustly determined even by considering only the UV luminosity density, local SMD, SN Ia and CCSN rates. However, considering the LGRB rate, along with an up-to-date, semi-empirical determination of the galaxy metallicity evolution throughout cosmic history, enables us to extend the study to LGRB progenitors and the cosmic metallicity evolution. Interestingly, an interplay emerges between the maximum metallicity of LGRB progenitors, and the dispersion in the galaxy metallicity distribution. We find an estimate of $Z_{max}=0.12^{+0.29}_{-0.05}\:Z_{\odot}$, that is in accordance with previous theoretical determinations in the literature. Moreover, we find that a fraction $N_{LGRB}=10^{+23}_{-7}\times 10^{-2}$ of stars in the mass range of LGRB progenitors actually produce a LGRB. This fraction agnostically accounts for all conditions required to have LGRB emission, including progenitor rotation, accretion disk mass, remnant nature, jet-launching and jet-energy conversion to LGRB. Moreover, it might contain the possible effect of LGRBs arising in binary systems.

Finally, we attempt at constraining the masses of CCSN and LGRB progenitors. Due to the simplicity of the model, and the limits of the adopted constraints, we are only able to obtain $M_{CC}^{up}\sim 58^{+37}_{-23}\:M_{\odot}$ for the upper mass limit of CCSN progenitors. Similarly, we get $M_{LGRB}^{low}<24\:M_{\odot}$, and $M_{LGRB}^{up}>51\:M_{\odot}$ for LGRB progenitors. We also test the hypothesis of an IMF varying with redshift, finding no evidence of evolution for the IMF slope, while the characteristic mass exhibits a mild increase with redshift. Despite the uncertainties associated to these results, they show the potential of this approach to study the progenitors of stellar transients and their IMF, especially considering the increased wealth of data in the future. Indeed, a number of missions relevant to this purpose are programmed or already ongoing, such as \textit{Euclid}, JWST, the \textit{Vera Rubin Observatory}, the \textit{Nancy Grace Roman Space Telescope}, the \textit{Zwicky Transient Facility}, \textit{WFIRST}, \textit{ULTIMATE-Subaru}, and the \textit{Gamow Explorer} (e.g. \cite{Refregier_2010,Laureijs_2011,Feindt_2019,Foley_2018,Foley_2019,Moriya_2019,Regos_2019,Wang_2019,Rose_2021,White_2021,Dessart_2022,Fryer_2022,Bailey_2023,Subrayan_2023,Arendse_2024,Bozzo_2024,Strotjohann_2024}). In particular, up to thousands or even tens of thousands SNe are expected to be observed, reaching out to higher redshifts than what currently set (e.g. \cite{Refregier_2010,Laureijs_2011,Feindt_2019,Foley_2019,Regos_2019,Rose_2021,Bailey_2023}). Moreover, new determinations of the SFRD and/or UV luminosity density are promised, especially at high redshifts, e.g. by JWST\footnote{https://www.stsci.edu/jwst/}. These achievements might also allow to shed light on topics such as the competing scenarios for LGRB progenitors, their putative association with those of SNe Ib/c, and the IMF universality.

\authorcontributions{Conceptualization: F.G., L.B., A.L.; methodology: F.G., L.B., A.L.; validation: L.B, A.L., G.G, O.S.S., R.S., M.S.; writing: F.G. All authors have read and agreed to the published version of the manuscript.}

\funding{This work is partially funded from the projects: "Data Science methods for MultiMessenger Astrophysics \& Multi-Survey Cosmology'' funded by the Italian Ministry of University and Research, Programmazione triennale 2021/2023 (DM n.2503 dd. 9 December 2019), Programma Congiunto Scuole; Italian Research Center on High Performance Computing Big Data and Quantum Computing (ICSC), project funded by European Union - NextGenerationEU - and National Recovery and Resilience Plan (NRRP) - Mission 4 Component 2 within the activities of Spoke 3 (Astrophysics and Cosmos Observations); PRIN MUR 2022 project n. 20224JR28W "Charting unexplored avenues in Dark Matter"; INAF Large Grant 2022 funding scheme with the project "MeerKAT and LOFAR Team up: a Unique Radio Window on Galaxy/AGN co-Evolution; INAF GO-GTO Normal 2023 funding scheme with the project "Serendipitous H-ATLAS-fields Observations of Radio Extragalactic Sources (SHORES)".}

\dataavailability{N/A} 

\acknowledgments{We thank A. Bressan for insightful discussions concerning the exploitation of the \texttt{PARSEC22} code, and P. Salucci for useful indications about galaxy stellar mass functions. We also thank T. Ronconi for precious advice in the implementation of the MCMC. Finally, we kindly thank F. Addari and C. Ugolini for useful conversations about type Ia and core-collapse supernovae.}

\conflictsofinterest{The authors declare no conflict of interest.} 

\renewcommand{\nomname}{Abbreviations}

\nomenclature{BH}{Black Hole}

\nomenclature{CCSN}{Core-Collapse Supernova}

\nomenclature{DD}{Double-Degenerate channel for type Ia supernovae}

\nomenclature{DTD}{Delay Time Distribution}

\nomenclature{FMR}{Fundamental Metallicity Relation}

\nomenclature{GSMF}{Galaxy Stellar Mass Function}

\nomenclature{IFMR}{Initial-Final Mass Relation}

\nomenclature{IGM}{InterGalactic Medium}

\nomenclature{IMF}{Initial-Mass Function}

\nomenclature{ISM}{InterStellar Medium}

\nomenclature{JWST}{James Webb Space Telescope}

\nomenclature{$\Lambda$CDM}{Lambda Cold Dark Matter cosmological model}

\nomenclature{LGRB}{Long Gamma-Ray Burst}

\nomenclature{MCMC}{Markov-Chain Monte Carlo}

\nomenclature{SB}{StarBurst Galaxy}

\nomenclature{SD}{Single-Degenerate channel for type Ia supernovae}

\nomenclature{SN Ia(Ib/c)}{type Ia(Ib/c) Supernova}

\nomenclature{SFR(D)}{Star Formation Rate (Density)}

\nomenclature{SMD}{Stellar Mass Density}

\nomenclature{SN}{Supernova}

\nomenclature{WD}{White Dwarf}

\nomenclature{(ZA)MS}{(Zero-Age) Main Sequence}

\printnomenclature


\appendixstart
\appendix

\section{Update on LGRB rate determination by Ghirlanda and Salvaterra 2022}\label{appendix:LGRB_update}

\begin{figure*}
    \includegraphics[width=0.7\textwidth]{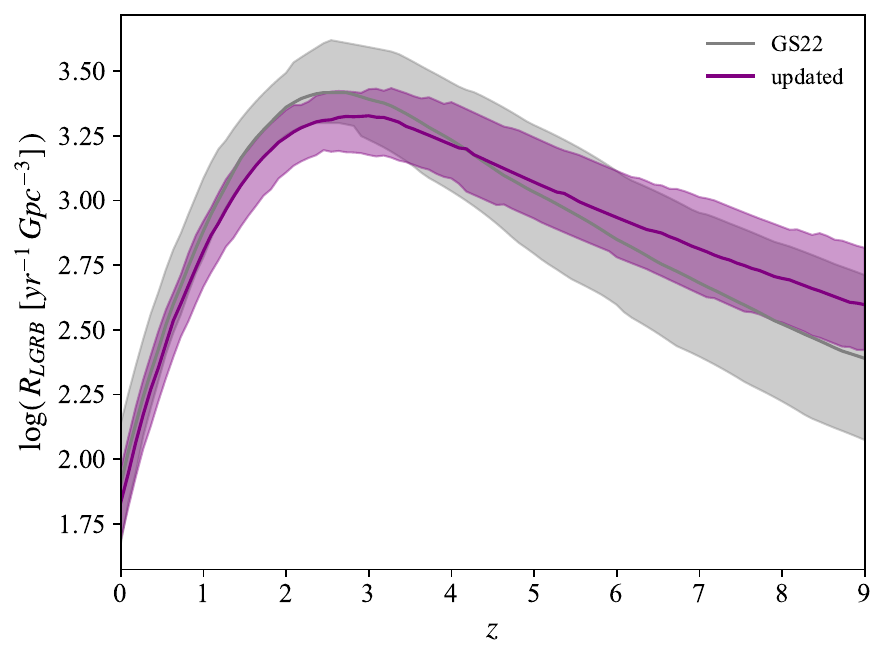}
    \caption{Comparison between the LGRB rate as a function of redshift obtained in \cite{Ghirlanda_2022} (grey line), with the updated determination employed in this work, accounting for more recent LGRB observations (purple line). Bands represent the corresponding $1\sigma$ uncertainties.}
    \label{fig:update_LGRB_comparison}
\end{figure*}

In this work, we consider an updated version of the LGRB rate determination as a function of redshift by \cite{Ghirlanda_2022}. This is obtained by updating the BAT6 sample used in \cite{Ghirlanda_2022}, which contained 79 GRBs up to May 2014, with all the bursts that satisfy the definition of the BAT6 sample as described in \cite{Salvaterra_2012}. This new sample contains 124 GRBs up to December 2022. With this new sample, the updated parameters (with their associated 1$\sigma$ uncertainties) describing the LGRB rate as a function of redshift (see Equation 1 of \cite{Ghirlanda_2022}) are: $\rho_{0}=68.33^{+23.87}_{-19.68}$ $Gpc^{-3}$ $yr^{-1}$,  $p_{z,1}=3.13^{+0.25}_{-0.28}$, $p_{z,2}=3.44^{+0.21}_{-0.25}$, $p_{z,3}=5.33^{+0.37}_{-0.33}$, and the luminosity evolution parameter is $\delta=1.05^{+0.38}_{-0.18}$. In Figure \ref{fig:update_LGRB_comparison}, we show a comparison between the two determinations. As one can see, the new LGRB rate peaks at slightly higher redshift, around $z\sim 3$, and decreases less steeply with respect to the old one, leading to an enhanced occurrence of LGRBs at high redshift.

\section{Additional MCMC results}\label{appendix:more_mcmc}

\begin{figure*}
    \includegraphics[width=\textwidth]{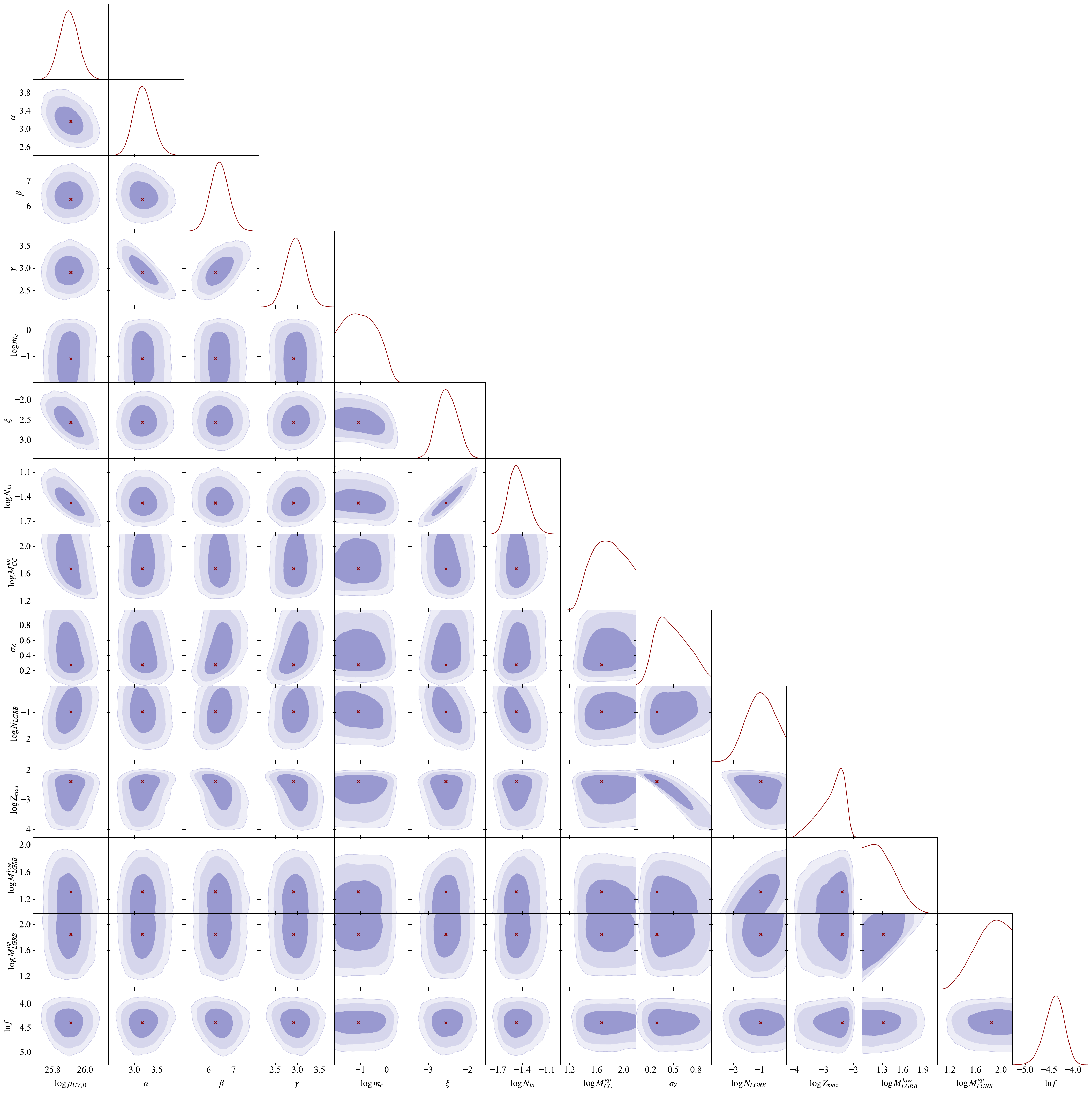}
    \caption{Same as Figure \ref{fig:MCMC_ONE}, for the whole parameter space.}
    \label{fig:MCMC_all}
\end{figure*}

We report in Figure \ref{fig:MCMC_all} the corner plot containing the posterior probabilities we obtain for the whole parameter space, regarding the IMF, $\rho_{UV}(z)$, $\rho_{\star,0}$, $R_{Ia}(z)$, $R_{CC}(z)$, $R_{LGRB}(z)$ and the cosmic $Z$ evolution.

\begin{figure*}
    \includegraphics[width=\textwidth]{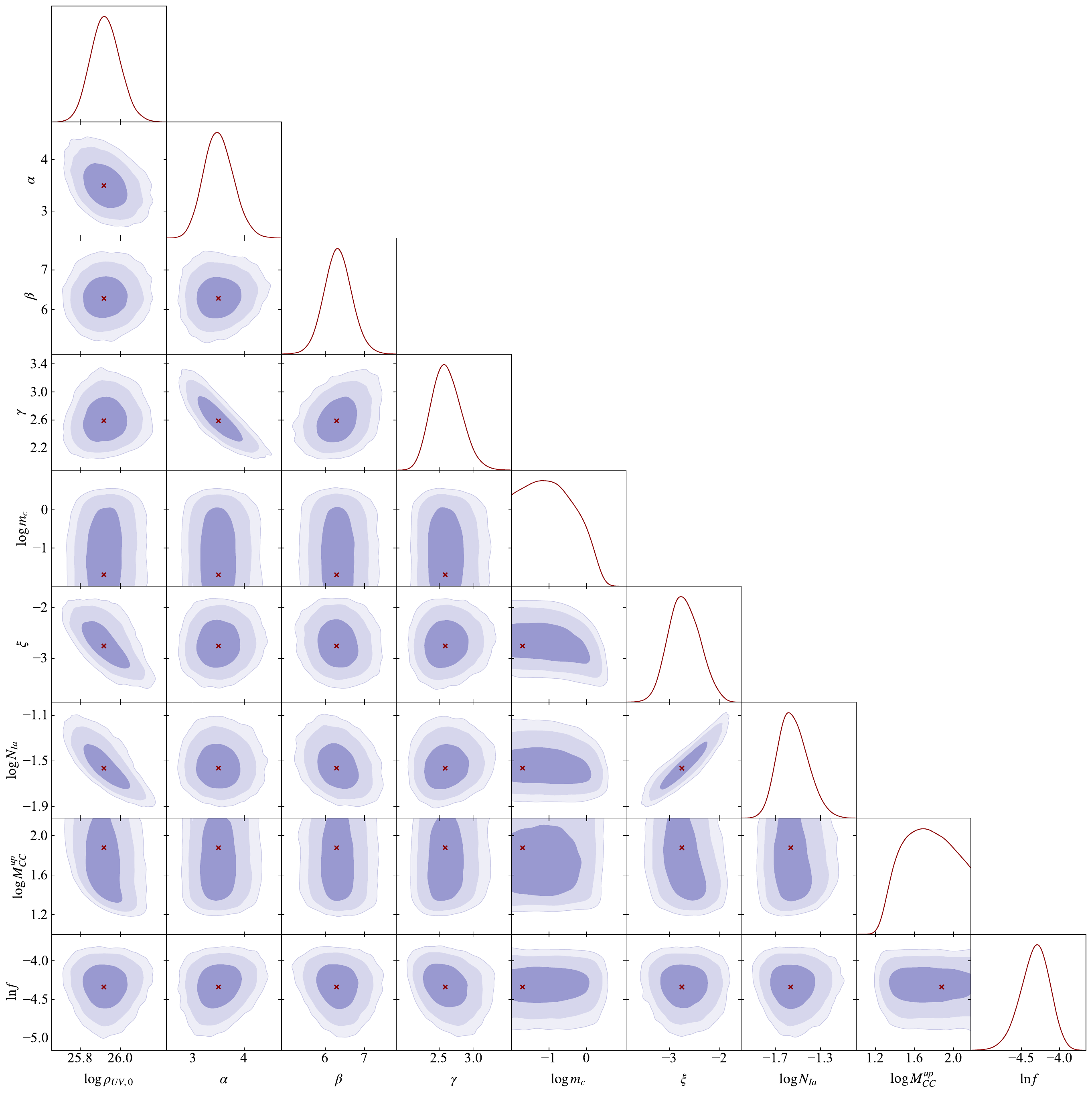}
    \caption{Same as Figure \ref{fig:MCMC_all}, for the case where we do not consider LGRBs and the cosmic $Z$ evolution.}
    \label{fig:MCMC_noLGRB}
\end{figure*}

In Figure \ref{fig:MCMC_noLGRB} and Table \ref{tab:fit_noLGRB}, we instead report the MCMC results obtained without considering LGRBs and the cosmic $Z$ evolution, only employing the constraints on $\rho_{UV}(z)$, $\rho_{\star,0}$, $R_{Ia}(z)$ and $R_{CC}(z)$. As one can see, these results are compatible with those shown in Figure \ref{fig:MCMC_all} (or \ref{fig:MCMC_ONE}) and Table \ref{tab:MCMC_all}.

\begin{table}
\centering
\captionof{table}{Estimates of the model parameters for the case where we do not consider LGRBs and the cosmic $Z$ evolution.}
\begin{tabular} { l c c }
\hline
{\boldmath$\log{\rho_{UV,0}}$} & $25.93^{+0.07}_{-0.08}  $ & $\log(erg\:s^{-1}\:Mpc^{-3})$\\
\vspace{0.5mm}
{\boldmath$\alpha         $} & $3.50^{+0.27}_{-0.31}      $ & - \\
\vspace{0.5mm}
{\boldmath$\beta          $} & $6.32\pm 0.36              $ & - \\
\vspace{0.5mm}
{\boldmath$\gamma         $} & $2.61^{+0.19}_{-0.23}      $ & - \\
\vspace{0.5mm}
{\boldmath$\log{m_c}      $} & $-0.95^{+0.39}_{-0.98}     $ & $\log(M_{\odot})$\\
\vspace{0.5mm}
{\boldmath$\xi            $} & $-2.71^{+0.30}_{-0.33}     $ & - \\
\vspace{0.5mm}
{\boldmath$\log{N_{Ia}}   $} & $-1.54^{+0.11}_{-0.15}     $ & - \\
\vspace{0.5mm}
{\boldmath$\log{M_{CC}^{up}}$} & $1.73\pm 0.24              $ & $\log(M_{\odot})$\\
\vspace{0.5mm}
{\boldmath$\ln{f}         $} & $-4.32^{+0.21}_{-0.17}     $ & - \\
\hline
\label{tab:fit_noLGRB}
\end{tabular}
\end{table}

\section{Additional results for IMF variation}\label{appendix:more_results_IMFvar}

\begin{figure*}
    \includegraphics[width=\textwidth]{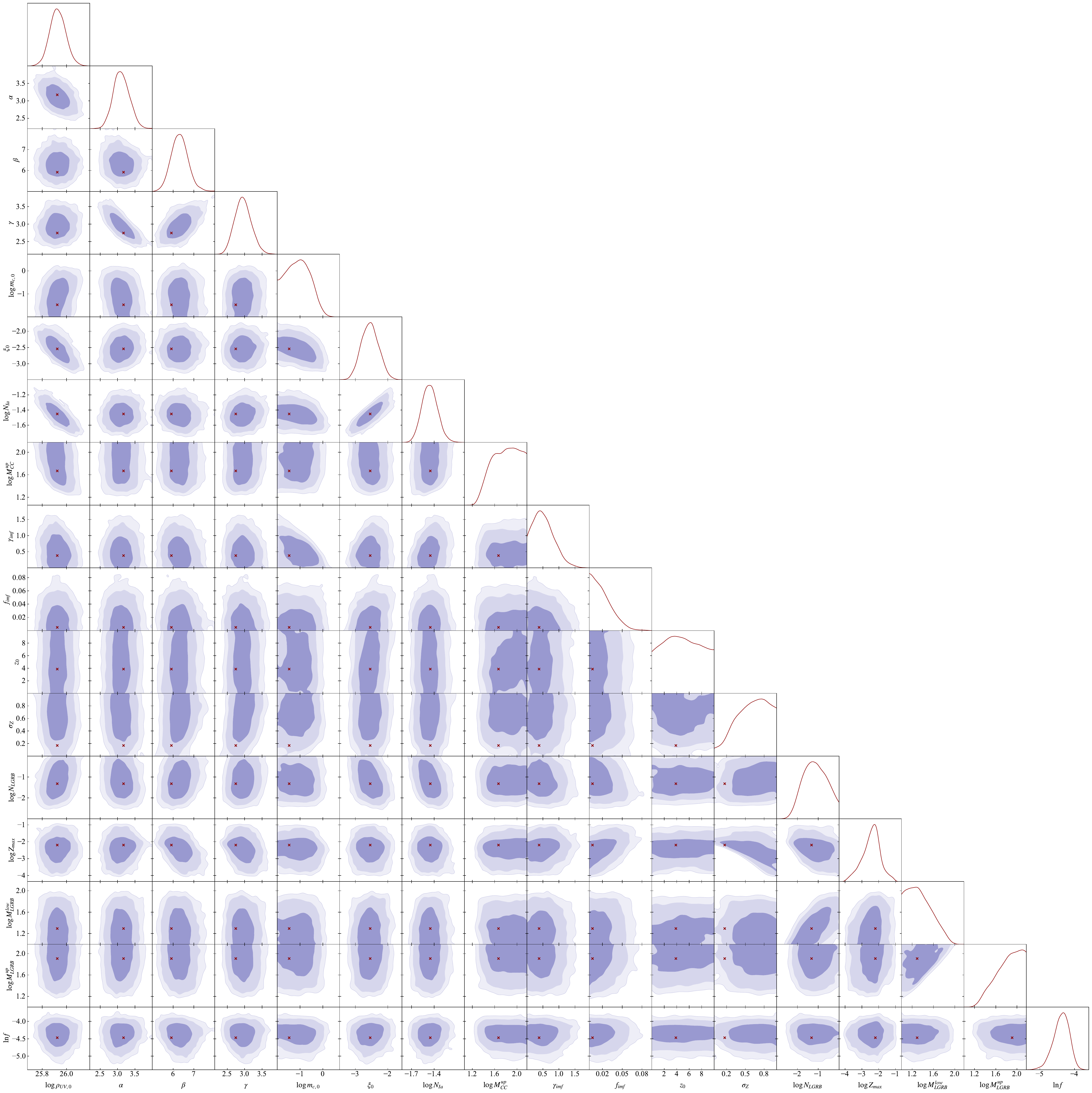}
    \caption{Corner plot with individual and joint posterior probabilities for the case of IMF variable with redshift, for the whole parameter space. For other details, see the caption of Figure \ref{fig:MCMC_ONE}.}
    \label{fig:MCMC_all_IMFvar}
\end{figure*}

In this section, we present the corner plot for the whole parameter space obtained in the case of IMF variable with redshift (Figure \ref{fig:MCMC_all_IMFvar}), as well as Table \ref{tab:MCMC_IMFvar} with all parameter estimates, to supplement the results shown in Section \ref{sec:IMF_var}.

\begin{table}
\centering
\captionof{table}{Estimates of the model parameters in the case of IMF variable with redshift.}
\begin{tabular} { l c c }
\hline
{\boldmath$\log{\rho_{UV,0}}$} & $25.93\pm 0.07           $ & $\log(erg\:s^{-1}\:Mpc^{-3})$\\
\vspace{0.5mm}
{\boldmath$\alpha         $} & $3.11^{+0.22}_{-0.25}      $ & - \\
\vspace{0.5mm}
{\boldmath$\beta          $} & $6.31^{+0.38}_{-0.42}      $ & - \\
\vspace{0.5mm}
{\boldmath$\gamma         $} & $2.95^{+0.22}_{-0.26}      $ & - \\
\vspace{0.5mm}
{\boldmath$\log{m_{c,0}}  $} & $-1.08\pm 0.51             $ & $\log(M_{\odot})$\\
\vspace{0.5mm}
{\boldmath$\xi_0          $} & $-2.55^{+0.25}_{-0.29}     $ & - \\
\vspace{0.5mm}
{\boldmath$\log{N_{Ia}}   $} & $-1.46^{+0.10}_{-0.11}   $ & - \\
\vspace{0.5mm}
{\boldmath$\log{M_{CC}^{up}}$} & $1.79^{+0.38}_{-0.16}      $ & $\log(M_{\odot})$\\
\vspace{0.5mm}
{\boldmath$\gamma_{imf}   $} & $0.53^{+0.21}_{-0.43}      $ & - \\
\vspace{0.5mm}
{\boldmath$f_{imf}        $} & $< 0.026                  $ & - \\
\vspace{0.5mm}
{\boldmath$z_0            $} & ---                          & - \\
\vspace{0.5mm}
{\boldmath$\sigma_Z       $} & $0.60^{+0.39}_{-0.12}      $ & - \\
\vspace{0.5mm}
{\boldmath$\log{N_{LGRB}} $} & $-1.15\pm 0.56             $ & - \\
\vspace{0.5mm}
{\boldmath$\log{Z_{max}}  $} & $-2.44^{+0.56}_{-0.46}     $ & - \\
\vspace{0.5mm}
{\boldmath$\log{M_{LGRB}^{low}}$} & $< 1.45                    $ & $\log(M_{\odot})$\\
\vspace{0.5mm}
{\boldmath$\log{M_{LGRB}^{up}}$} & $> 1.73                    $ & $\log(M_{\odot})$\\
\vspace{0.5mm}
{\boldmath$\ln{f}         $} & $-4.38^{+0.26}_{-0.19}     $ & - \\
\hline
\label{tab:fit_IMFvar_all}
\end{tabular}
\end{table}

\section{Parameter estimates for the FMR variation}\label{appendix:FMR_variations}

We report in Table \ref{tab:FMR_Curti20} the parameter estimates returned by the MCMC for the FMR variation by \cite{Curti_2020}, without the offset from \cite{Curti_2023} that we instead implement in this work.

\begin{table}
\centering
\captionof{table}{Parameter estimates for the variation with FMR by \cite{Curti_2020}, without the offset from \cite{Curti_2023}.}
\begin{tabular} { l c c }
\hline
{\boldmath$\log{\rho_{UV,0}}$} & $25.88^{+0.05}_{-0.06}  $ & $\log(erg\:s^{-1}\:Mpc^{-3})$\\
\vspace{0.5mm}
{\boldmath$\alpha         $} & $3.32\pm 0.23              $ & - \\
\vspace{0.5mm}
{\boldmath$\beta          $} & $5.93\pm 0.28              $ & - \\
\vspace{0.5mm}
{\boldmath$\gamma         $} & $2.74\pm 0.20              $ & - \\
\vspace{0.5mm}
{\boldmath$\log{m_c}      $} & $-0.94^{+0.60}_{-0.74}     $ & $\log(M_{\odot})$\\
\vspace{0.5mm}
{\boldmath$\xi            $} & $-2.46\pm 0.25             $ & - \\
\vspace{0.5mm}
{\boldmath$\log{N_{Ia}}   $} & $-1.41^{+0.11}_{-0.13}     $ & - \\
\vspace{0.5mm}
{\boldmath$\log{M_{CC}^{up}}$} & $1.76\pm 0.22              $ & $\log(M_{\odot})$\\
\vspace{0.5mm}
{\boldmath$\sigma_Z       $} & $0.36^{+0.11}_{-0.14}      $ & - \\
\vspace{0.5mm}
{\boldmath$\log{N_{LGRB}} $} & $-0.69^{+0.68}_{-0.18}     $ & - \\
\vspace{0.5mm}
{\boldmath$\log{Z_{max}}  $} & $-2.76^{+0.46}_{-0.30}     $ & - \\
\vspace{0.5mm}
{\boldmath$\log{M_{LGRB}^{low}}$} & $< 1.34                    $ & $\log(M_{\odot})$\\
\vspace{0.5mm}
{\boldmath$\log{M_{LGRB}^{up}}$} & $1.82^{+0.33}_{-0.13}      $ & $\log(M_{\odot})$\\
\vspace{0.5mm}
{\boldmath$\ln{f}         $} & $-4.32^{+0.19}_{-0.16}     $ & - \\
\hline
\label{tab:FMR_Curti20}
\end{tabular}
\end{table}

\begin{adjustwidth}{-\extralength}{0cm}

\reftitle{References}

\bibliography{bib}

\PublishersNote{}
\end{adjustwidth}
\end{document}